\documentclass[fleqn,usenatbib,linenumbers]{mnras}

\usepackage{url}

\usepackage{newtxtext,newtxmath}

\usepackage[export]{adjustbox}
\usepackage{subfig}

\usepackage[T1]{fontenc}

\DeclareRobustCommand{\VAN}[3]{#2}
\let\VANthebibliography\thebibliography
\def\thebibliography{\DeclareRobustCommand{\VAN}[3]{##3}\VANthebibliography}

\usepackage{graphicx}	
\usepackage{amsmath}	
\usepackage{CJK}
\usepackage{pgfplotstable}
\usepackage{threeparttable}
\usepackage{changepage}
\usepackage{csvsimple}

\newcommand\gray{$\gamma$-ray}
\newcommand\Edot{$\dot{E}$}
\newcommand\Pdot{$\dot{P}$}
\newcommand\FluxUnit{ph cm$^{-2}$ s$^{-1}$}

\graphicspath{{./}{figures/}}
\pgfplotsset{compat=1.17}

\title[Stacking of Gamma-ray Pulsars]{A Stacking Survey of Gamma-ray Pulsars}

\author[Y. Song et al.]{
Yuzhe Song(宋宇哲),$^{1,2,3,4,5}$\thanks{E-mail: yuzhesong@swin.edu.au (YS)}
Timothy A. D. Paglione,$^{1,2,3}$
Joshua Tan$^{3,6}$,
Charles Lee-Georgescu$^{6}$
\newauthor and Danisbel Herrera$^{3,7}$
\\
$^{1}$Physics Program, the Graduate Center, City University of New York, 365 Fifth Ave., New York, NY 10016, USA\\
$^{2}$Department of Earth \& Physical Sciences, York
College, City University of New York, 94-20 Guy R. Brewer Blvd.,
Jamaica, NY 11451, USA\\
$^{3}$Department of Astrophysics, American Museum of
Natural History, Central Park West at 79th Street, New York, NY 10024, USA\\
$^{4}$Center for Astrophysics and Supercomputing, Swinburne University of Technology, John St, Hawthorn, VIC 3122, Australia\\
$^{5}$OzGrav, ARC Center for Excellence in Gravitational Wave Discovery\\
$^{6}$Department of Natural Sciences, LaGuardia Community College, City University of New York, 31-10 Thomson Ave, Long Island City, NY 11101, USA\\
$^{7}$Department of Physics, New York University, 726 Broadway, New York, NY 10003, USA}

\date{Accepted 2023 July 17. Received 2023 July 15; in original form 2023 May 15}

\pubyear{2023}

\begin{document}
\begin{CJK*}{UTF8}{gbsn}
\label{firstpage}
\pagerange{\pageref{firstpage}--\pageref{lastpage}}
\maketitle

\begin{abstract}
We report on a likelihood stacking search for gamma-ray pulsars at 362 high-latitude locations that coincide with known radio pulsar positions. We observe a stacked signal conservatively 2.5$\sigma$ over the background. Stacking their likelihood profiles in spectral parameter space implies a pulsar-like spectral index and a characteristic flux
a factor of 2 below the Fermi-LAT point source sensitivity, assuming isotropic/unbeamed emission from all sample pulsars. The same procedures performed on empty control fields indicate that the pulsars as a population can be distinguished from the background with a $\Delta$(TS) = 28 at the peak location (or 4.8$\sigma$), and the stacked spectra of the control fields are distinctly softer than those of the pulsars. This study also probes a unique region of parameter space populated by low \Edot\ pulsars, most of which have no \gray\ ephemeris available, and is sensitive to high duty cycles. We also discusses the possible \gray\ emission mechanism from such pulsars.

\end{abstract}

\begin{keywords}
pulsars: general -- stars: neutron -- gamma-rays: stars -- gamma-rays: diffuse background -- methods: data analysis -- methods: statistical
\end{keywords}



\section{Introduction}

Identifying astrophysical $\gamma$-ray sources is one of the main goals of the Fermi Gamma-ray Space Telescope. While the updated 10-year source catalogue, 4FGL-DR2 \citep[the 4FGL-DR2,][]{ajello2020_4FGL}, hereafter the 4FGL, identifies 279 sources as pulsars, and the 12-year catalogue, 4FGL-DR3, identifies 30 more \gray\ pulsars \citep{dr3}, this is less than 10\% of the pulsars in the Australia National Telescope Facility (ATNF) pulsar catalogue \citep{atnf}. 

With over fiftean years of almost continual all-sky survey data, the Fermi Large Area Telescope (LAT) is now becoming sensitive to pulsars that are below the putative \gray\ pulsar ``death line'' in spin-down luminosity \Edot\ $\lesssim 10^{33}\ \rm{erg}\ \rm{s}^{-1}$ \citep{smith2019} where the majority of rotationally-powered pulsars reside. Studies of faint \gray\ pulsars \citep[e.~g.,][]{hou2014, bruel2019} illustrate both the potential and the difficulties of pushing the limits of sensitivity. Bright backgrounds, steep source spectra with low energy cutoffs, and broad, hard-to-discern pulses make identifying low-luminosity pulsars challenging. Nevertheless, it is worth trying to detect this population because they are found in a corner of pulsar parameter space predicted to probe rare alignment geometries and potentially novel emission mechanisms \citep{smith2019}.

Stacking methods in gamma-ray observations date back to the operational era of the Compton Gamma Ray Observatory's Energetic Gamma Ray Experiment Telescope (EGRET) when stacking studies on galaxy clusters \citep{reimer2003} and galaxies \citep{cillis2004, cillis2005, cillis2007} were carried out. Other gamma-ray observatories like the High Altitude Water Cherenkov (HAWC) gamma-ray observatory used a similar joint-likelihood analysis \citep{hawc2021} that resembles the stacking techniques developed for EGRET and Fermi-LAT.  A significant amount of \gray\ flux from the known pulsars not yet detected in \gray s should be present in the Fermi-LAT data. Authors such as \citet{huber2012} proposed it should be discoverable using stacking techniques, but such techniques have up until now been unsuccessful in probing this population \citep{mccann2015}. In contrast, stacking techniques that aggregate photon counts have been successfully employed in surveys of galaxy clusters \citep{dutson2013, griffin2014, prokhorov2014, reiss2018}, providing strong upper limits or detections of galaxy clusters as a population. More recently, an improved likelihood-stacking technique was developed and successfully characterized the faint blazar and starburst galaxy populations at GeV energies \citep{ajello2020_sbgs, paliya2019}, pushing well beyond the point-source detection limit of the LAT. This study extends these kinds of likelihood-stacking methods to the pulsar population with a new approach that stacks the measured Test Statistic (TS) values for pulsar fields, and, in so doing, provides some of the first measurements of this undetected population. This new method quantifies the significance of a stack observation and, when applied while varying spectral parameters, characterizes the \gray\ emission of the pulsar population under investigation. Such information can be used to inform how undetected populations of pulsars contribute to the diffuse \gray\ background, the \gray\ excess observed at the Galactic center, and \gray\ flux from globular clusters. 

With this project, we show evidence for the existence of a population of sub-threshold \gray\ pulsars and begin the investigation of this elusive population. We probe pulsars with high duty cycles with greatly improved sensitivity of the stacking methods. Furthermore, without assuming any beaming geometry of the sample pulsars, 
this work can potentially characterize un-pulsed \gray\ emission that may be produced by this population. The organization of this paper is as follows: In \S~\ref{sec:obs}, we describe our list of targets, the observational methods, and the stacking techniques. We outline the results of our analysis in \S~\ref{sec:results}, where we summarize the new candidate population of sub-threshold pulsars including information that can be extracted from our stacking analysis. We also analyze control fields to validate the results. The overall properties of these sources are placed in context with the known pulsar population in \S~\ref{sec:discussion}. This is followed by a brief conclusion and five short appendices which detail justifications for the specific choices in the methodology detailed in the main text and additional analysis of particularly significant candidate sources.

\section{Observations and Methods}
\label{sec:obs}
We identify a sample of undetected pulsars in relatively uncrowded parts of the \gray\ sky to perform our stacking analysis. These locations are then analyzed using specialized software for Fermi LAT observations and the results are stacked as a sum of the Test Statistic (TS, defined in \S~\ref{subsec:data_analysis}), which is derived from the likelihood of a source detection at a given location. Normal thresholds for detection (nominally TS > 25 as seen in the 4FGL) do not apply in such circumstances as, on average, the Fermi data tend to yield slightly positive TS values indicating that any given random point in the sky is somewhat favored to be modeled as having a source. To better understand the significance of a TS stack, a comparison to control fields---locations on the sky randomly selected to be similarly distributed as our test sample---is provided. 

\subsection{Data Selection}\label{subsec:dataselection}
We choose target pulsars from the ATNF pulsar catalogue version 1.64, November 2020\footnote{\raggedright \url{https://www.atnf.csiro.au/research/pulsar/psrcat/}} that were not in the 4FGL catalogue using the {\tt psrqpy} Python package\footnote{\raggedright \url{https://psrqpy.readthedocs.io/en/latest/}} \citep{pitkin2018}. To optimize sensitivity, we avoid pulsars located in regions with complicated $\gamma$-ray emission backgrounds, such as the Galactic plane and bulge, and regions near extremely bright $\gamma$-ray sources. We therefore limit the search to high latitudes ($|b| > 20 \degr$). Detailed reasoning for selecting this latitude cut is described in Appendix~\ref{appendix:latcut}. We also avoid all globular clusters and the Magellanic Clouds. A total of 362 pulsar positions met these initial selection criteria\footnote{the complete list of pulsars in this study can be found at: \url{https://github.com/yuzhesong/stacking\_funcs\_tutorial/blob/main/pulsars\_results\_2023.csv}}.

\begin{figure*}
\centering
\includegraphics[width=0.8\textwidth]{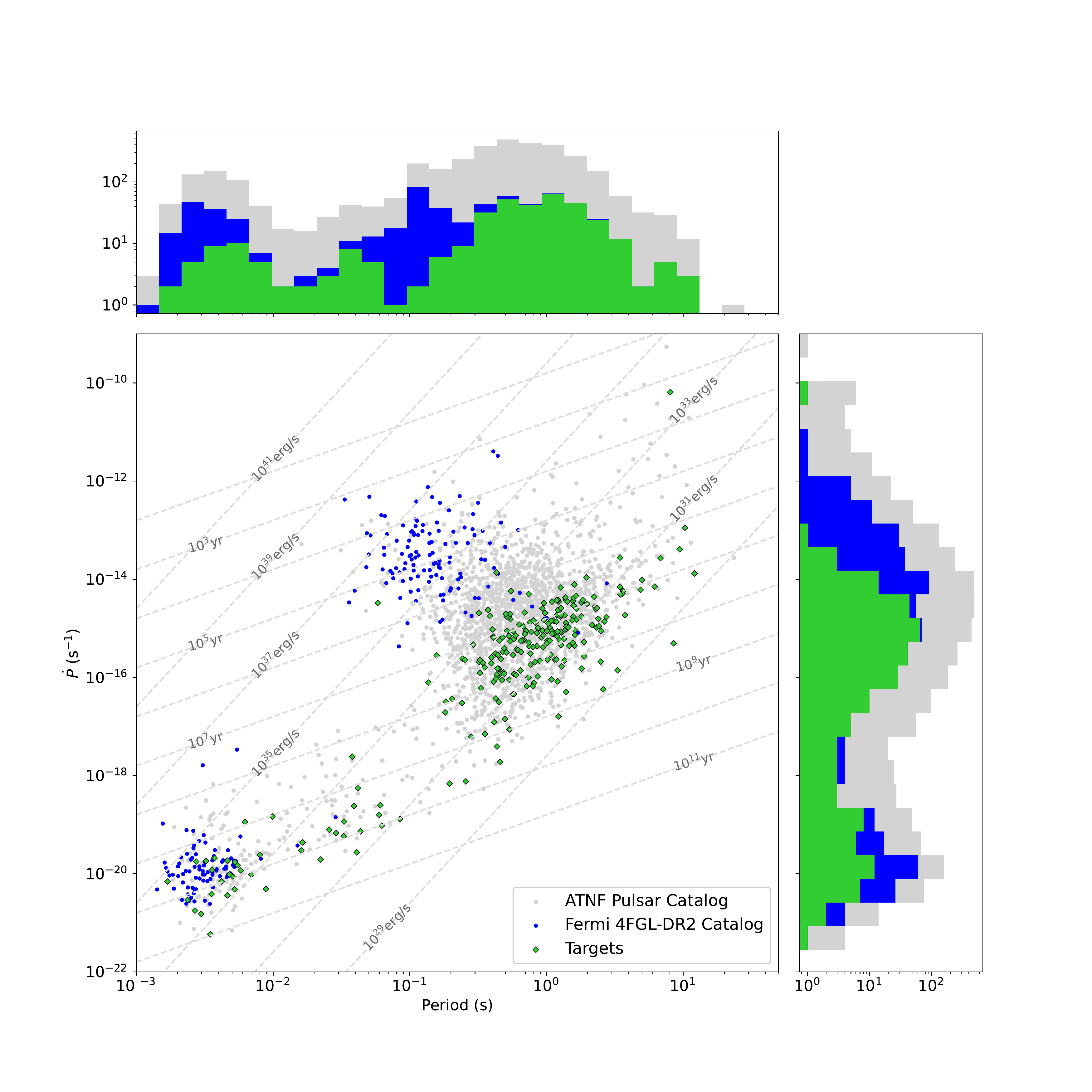}
    \caption{$P$-$\dot{P}$ diagram of the pulsars surveyed: sub-threshold sources that were stacked (green diamonds), Fermi catalogue pulsars (blue dots) and ATNF pulsars (gray dots). Lines of constant characteristic age and \Edot\ are indicated. The right side histogram shows the distribution of $\dot{P}$ for all the pulsars for which this has been measured; the top panel histogram shows the distribution of $P$ for all the pulsars including those not appearing on the scatterplot due to a lack of $\dot{P}$ measurements. The histograms have the same color coding as the symbols, and the color bars are placed on top of each category. Out of the entire pulsar population, 482 pulsars do not have \Pdot\ measurements and 26 of these do not have period measurements.} 
    \label{fig:ppdot_all}
\end{figure*}

Our target ATNF pulsars---along with the full ATNF catalogue and 4FGL pulsars---are plotted in Fig.~\ref{fig:ppdot_all}. This high latitude study, while aimed primarily at reducing the background systematics, also probes a unique region of parameter space populated by older, transitional, and recycled millisecond pulsars (MSPs). The surveyed source distribution also more closely reflects the overall distribution of objects in the ATNF compared to those in the 4FGL. 

Over twelve years of data from the LAT were used in this survey. The third revision of the PASS8 data, P8R3, released on Nov 26, 2018 was used along with the 4FGL and diffuse background models \citep{abdo2009}. Data analysis in this work is performed with {\tt Fermipy} \citep{fermipy} version 1.0.1\footnote{\url{https://fermipy.readthedocs.io/en/latest/}} based on the Fermi Science tools Anaconda distribution \citep{anaconda2020} version 2.0.8\footnote{\url{https://fermi.gsfc.nasa.gov/ssc/data/analysis/}}. Data were chosen between Mission Elapsed Time (MET) 239560000s (MJD = 54682) and 641950000s (MJD = 59340). 
Each region of interest (ROI) is a $21.2\degr \times 21.2\degr$ square, which corresponds to a $\sim 15\degr$ radius ROI.
The data were also filtered using a zenith angle cut of $90\degr$ to avoid bright emission from the Earth. Good time intervals were chosen with  conditions {\tt DATA\_QUAL==1 \&\& LAT\_CONFIG==1}. The data were binned in 30 logarithmically-spaced energy bins (unbinned analyses for such a large duration of time are computationally prohibitive). An all-sky livetime cube and all-sky exposure cube\footnote{\url{https://fermi.gsfc.nasa.gov/ssc/data/analysis/scitools/binned\_likelihood\_tutorial.html}} within the specified time range from above were created and utilized for all ROIs.

Some previous studies have uncovered potential systematic effects studying faint populations at low energies \citep{paliya2020, principe2021}. We took a data-based approach to quantify this effect by conducting the likelihood analysis in two different energy ranges. Our primary stacking analysis only uses photons in the energy range of 300 MeV to 100 GeV to constrain the spectral energy distributions (SEDs) and calculate final fluxes. An analysis that includes additional photon data from 100 MeV to 300 MeV is found in Appendix~\ref{appendix:100MeV}. We also note that raising this lower energy limit further, e.g., to 1 GeV, would improve the differentiation of target sources and the background even more, but at too great a cost. The average cutoff energy of all Fermi detected pulsars is 830 MeV, so such a restriction eliminates most of the \gray\ signal we are attempting to uncover.

We also have created a library of 540 control fields from which we pull samples to validate the likelihood results, identify any systematic effects, and estimate the background signal. The control fields have the same radius of $15\degr$ and are drawn from locations within the same Galactic latitude and longitude distributions as the target pulsars. The center of each control field is chosen to be at least $1\degr$ away from any known 4FGL source to be more certain that any measured signal can be attributed solely to fluctuations in the background.

\subsection{Fermi-LAT Data Analysis}\label{subsec:data_analysis}
{\tt Fermipy} automatically reads the currently available version of the 4FGL catalogue and creates a model file for each ROI, along with the Galactic diffuse background emission and the isotropic background. We model an additional source at the center of each ROI as a power law with an exponential cutoff (PLEC) as the standard \gray\ spectral shape for pulsars \citep{2PC}. For the model parameters, {\tt prefactor}, {\tt index1}, and {\tt cutoff} are set free to be fitted, while {\tt scale} is set to 1000 MeV and {\tt index2} is set to 1.0 throughout the analysis. We set free the spectral parameters of all sources within $7.5\degr$ from the ROI center, and the diffuse background models. The pulsar positions are fixed at the values reported in ATNF catalog. We note that some of the pulsars in the sample do not have a timing position and may have relatively uncertain coordinates. However, it is beyond the scope of this work to provide precise locations of these pulsars, hence we fix the coordinates of the pulsars throughout our analysis.
This process yields the TS value of each pulsar, defined as TS $=2 \log{{\cal L}/{\cal L}_0}$, a comparison of the log likelihoods of the source model and the null hypothesis (no source, ${\cal L}_0$). Usually, $\sqrt{TS}$ is used as a rough estimate of the source significance. 

Additional unmodeled sources with TS $> 25$ were identified using the {\tt find\_sources} function in {\tt Fermipy}, assuming they are power law sources.  

\subsection{Stacking Methods}
\label{subsec:stacking_method}
In \gray\ observations, counts maps of each ROI can be overlaid and added up to create a stacked counts map. Broadly speaking, the resulting ``stack'' can be processed through the likelihood analysis to estimate the signal significance. Stacking photon counts has been applied to varied populations such as galaxies, galaxy clusters, and flare stars \citep{cillis2004, cillis2005, cillis2007, griffin2014, song2020}. \citet{huber2012} was among the early efforts to stack Fermi-LAT data by comparing stacked signal and background counts to calculate their likelihoods and the TS value of any detection.
Direct likelihood stacking has also been conducted on blazars \citep{paliya2019, paliya2020}, star-forming galaxies \citep{ajello2020_sbgs}, and young radio galaxies and quasars \citep{principe2021}. These studies not only obtained sensitive upper limits on the \gray\ emission, but have also made robust detections of some populations, plus a characterization of their optimal average spectral parameters. This study takes these methods and extends them to the undetected pulsar population described in \S~\ref{subsec:dataselection} while adopting a new interpretation of the cumulative TS stacks for characterizing the significance of a population's potential signal over background and the TS maps of sources in spectral parameter space. 

To test the existence of an undetected population, the first method used in this study is to consider simply adding up the TS values of all the sources of interest (SOIs) and comparing that sum to the cumulative TS of a population of control field test sources. Here we assume that a TS stack of SOIs, when compared to a TS stack of ROIs of blank sky positions, will result in a higher cumulative TS value if there is an underlying population of \gray\ sources emitting below the detection threshold. 
Stacking TS values is a much more sensitive measure than simple counts stacks, which are overwhelmed by noise given the low counts.  
The calculation proceeds as follows: For $N$ sources, each with a TS value TS$_k = 2 \times ( \ln{{\cal L}}_{k, \text{source}} - \ln{{\cal L}}_{k, \text{null}} )$, the cumulative TS value is 
\begin{equation}
    {\text{TS}} = \sum_{k=1}^N \text{TS}_k = 2 \times \sum_{k=1}^N \ln{{\cal L}}_{k, \text{SOI}} - 2 \times \sum_{k=1}^N \ln{{\cal L}}_{k, \text{null}}
    \label{eq:cumTS}
\end{equation}
\noindent which is a comparison between the collective maximum source likelihood and the collective null likelihood. Similarly, the cumulative TS value for $N$ control field test sources is 
\begin{equation}
    {\text{TS}}_{\rm cf} = \sum_{k=1}^N \text{TS}_{k,\text{cf}} = 2 \times \sum_{k=1}^N \ln{{\cal L}}_{k, \text{cf}} - 2 \times \sum_{k=1}^N \ln{{\cal L}}_{k, \text{null}}.
    \label{eq:cumTS_cf}
\end{equation}

\noindent Subtracting Eq.~\ref{eq:cumTS_cf} from Eq.~\ref{eq:cumTS} effectively cancels the summed null term given appropriately chosen control fields. Thus, we can quantify the significance of the stack by introducing $\Delta$(TS)
\begin{equation}
    \Delta (\text{TS}) = 2 \times \sum_{k=1}^N \ln{{\cal L}}_{k, \text{SOI}} - 2 \times \sum_{k=1}^N \ln{{\cal L}}_{k, \text{cf}}
\end{equation}

\noindent If the stacked SOI TS values are consistently higher than the control field test source TS values, we may conclude that a signal exists, and the $\Delta$(TS) can be used to estimate the significance of the population, considering the control field likelihood term as the null. 

One complication to address is that there is considerable variation in the distribution of TS values at various locations. Thus, a naive $\chi^2$ interpretation of the $\Delta$(TS) value may overstate the significance of a stack's signal (see the left panel of Fig.~\ref{fig:results_1} for the distribution of TS values for SOIs and control field ROIs and how they differ from the $\chi^2$ distribution). To account for this potentiality, a bootstrap resampling scheme is adopted to provide a conservative estimate of the uncertainties. Given $N$ sources in either population in the stack, we randomly resample a given population $N$ times, and calculate their cumulative TS distributions as a function of the stacked number of ROIs. This bootstrapping procedure is repeated $M$ times, and the average and standard deviations of the cumulative TS values are calculated.

The entire process described above---a random sampling of $N$ sources $M$ times---is repeated for a total of $P$ realizations. This process returns the distribution of the mean final cumulative (stacked) TS values for the population. Multiple sampling of a population that exceeds $N$ (as in the case of our control fields) also yields a measure of the distribution for the final point of the stack. If a stack of sources is detected, the distributions of the cumulative TS value of the control fields should be statistically distinguishable from the final cumulative TS value of the source stack. 

In the right panel of Fig.~\ref{fig:results_1}, we quantify the significance of this signal by measuring the separation of the two stacks. For the target pulsars, the uncertainty of the stack is estimated by stacking randomly re-ordered ROIs which creates a characteristic spindle-shaped envelope (shown in blue) as the variance increases halfway through the stack and then decreases until the final cumulative TS value. In contrast, for control fields (shown in green), the bootstrap resampling method described above yields an uncertainty that increases with the number in the stack. The $\Delta(\textrm{TS})$ between the final stacks can be used to roughly estimate the separation between the two populations, and $\sqrt{\Delta({\rm TS})}$ between the bootstrap uncertainty of the control fields can be used to estimate the significance of the separation.

As a second method, we investigate the spectral properties of this pulsar population by performing a TS stack in spectral parameter space similar to the likelihood stack in \citet{paliya2019} and other works mentioned above. A TS map of the pulsar in spectral parameter space is created as described below. A point source with a PLEC spectrum is placed at the location of each pulsar fixing flux, spectral index, and cutoff energy. Only the background isotropic and Galactic diffuse normalizations are free to be fit by {\tt Fermipy}'s {\tt fit} function. This process returns a log-likelihood for the ROI given these spectral parameters for the central source. We repeat this process over a grid of flux and index values, keeping the cutoff energy fixed at 823.3 MeV to lower the degrees of freedom to 2. This cutoff energy is the median of all pulsars recorded in 4FGL-DR2. A detailed justification for fixing the cutoff energy is given in Appendix~\ref{appendix:3dparstack}. TS maps in flux-index parameter space for each ROI are made by subtracting the log likelihood of the grid point representing the null, which is chosen to be at the lowest flux ($7.2\times 10^{-12}$ \FluxUnit) and spectral index of $-4$, then multiplying this result by 2. The stacked TS map is then made by summing the individual TS maps. An over abundant amount of control fields is utilized for this study. To obtain the control field stack, we randomly choose the same number of control fields as the pulsar ROIs and add up all their TS maps, repeating this 1000 times, and averaging. Validation of this method is done by performing the analysis on a pulsar from the sample list, as described in \S~\ref{subsec:validation}.

\section{Results}
\label{sec:results}

\subsection{Cumulative TS Stack}
\label{subsec:preresults}

\begin{figure*}
\centering
\includegraphics[width=\textwidth]{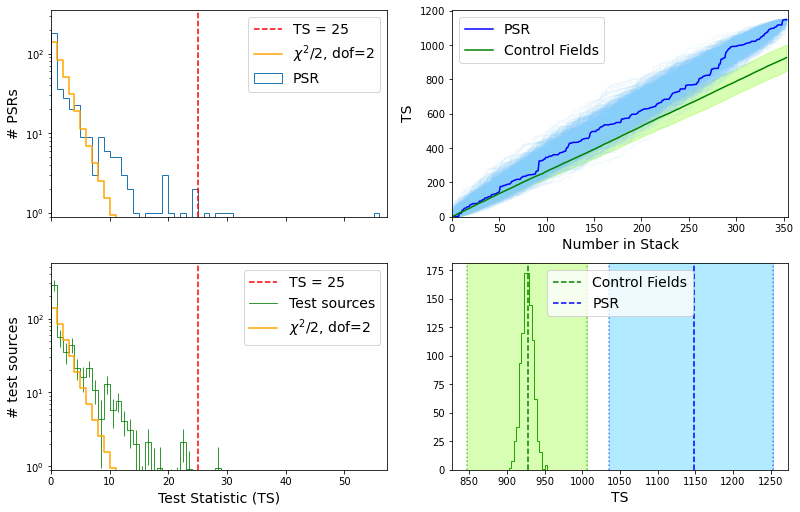}
\caption{Results of TS stacking. Top left: TS values of 362 target pulsars from analysis between 300 MeV and 100 GeV. Bottom left: TS values of 362 randomly sampled control field test sources. The histogram values and the error bars are the averages and standard deviations after 100 samplings of the 540 control fields. The orange histograms represent the $\chi^2$/2 distribution for 2 degrees of freedom. The red dashed vertical line represents TS $= 25$, the usual detection threshold. Top right: Cumulative TS values of the target pulsars (blue) and control field test sources (green). The light blue shaded area represents randomly reordered stacks of the pulsar TS values. The light green shaded areas represent uncertainties of the stack for control fields estimated from bootstrap resampling of the stacks. Bottom right: The final mean cumulative TS value for pulsars (vertical blue line) and control fields (green histogram). The dashed green line represents the mean of the distribution. The dotted lines represent the mean of the standard deviations of the 1000 realizations of the bootstrap resampling process.
\label{fig:results_1}}
    
\end{figure*}

Fig.~\ref{fig:results_1} shows the TS distributions of the central sources in the target ROIs measured above 300 MeV. The likelihood analyses for 9 of the pulsars in our sample did not converge and were omitted here. For comparison, the TS values for test sources at the centers of the control fields are also shown. The $\chi^2/2$ distribution for three degrees of freedom is also shown, which should be equivalent to the theoretical null \citep{mattox1996}. Some of the pulsars have TS values above 25 and are further discussed for the possibility of detection in Appendix~\ref{appendix:discussion}. 

To evaluate the uncertainty of the TS stacks, we did a bootstrap resampling of all ROIs $M=100$ times for a total of $P=1000$ realizations. The average cumulative TS value of the control fields is shown in two panels on the right of Fig.~\ref{fig:results_1} along with their standard deviations. The cumulative TS value for the pulsars, with a standard deviation estimated from the bootstrap resampling technique are shown in the same panels. There is a statistically significant difference between the cumulative TS values of the pulsars and the control field test sources, $\Delta$TS $>200$. 

We quantify the significance of the stacked signal in a few different ways. Firstly and most intuitively, a Kolmogorov-Smirnov test \citep{kolmogorov, smirnov} comparing ten thousand bootstrap resamplings of each distribution results in a statistic of $D=0.67$, verifying that the two distributions are distinct with near statistical certainty. Secondly, we compare the final cumulative TS values of the pulsars and control field test sources. For the control fields, the final cumulative TS value is around 940 following the bootstrapping procedures. The mean bootstrap uncertainty is 82 and is represented by the green shaded area in Fig.~\ref{fig:results_1}. For the pulsars, the final cumulative TS value is 1144. The mean bootstrap uncertainty is 108 and is represented with the blue shaded area in Fig.~\ref{fig:results_1}. Between the two populations, $\Delta$(TS) = 204, naively indicating a $14.3\sigma$ difference. More conservatively, using the bootstrap uncertainty ranges yields a $2.5\sigma$ difference. 

\subsection{Parameter Space Stacking Results}
\label{subsec:parstack_results}
\begin{figure}
    \centering
    \includegraphics[width=\columnwidth]{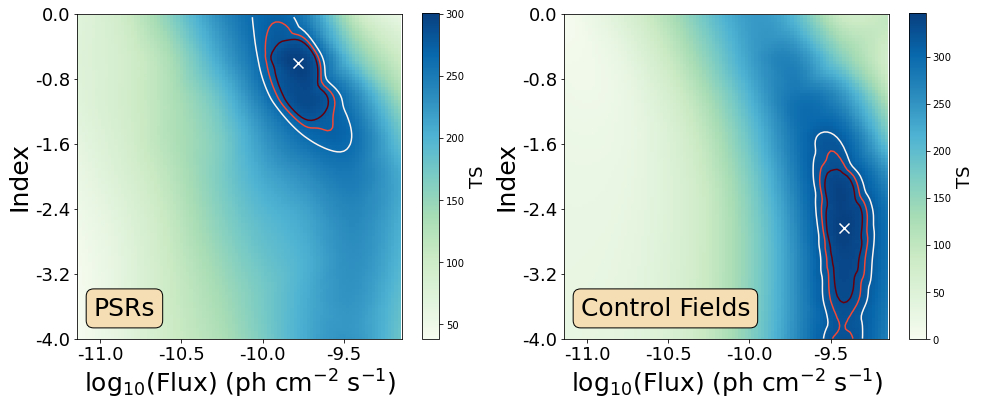}
    \caption{Parameter space stacks of pulsars and control fields with various weights. Left: parameter space stack of 353 ROIs with candidate pulsars. The white X marks the optimal parameters with the largest TS value of 300. Right: Parameter space stack for control fields test sources, produced by randomly choosing 353 ROIs out of 540 control fields and repeating this random selection 1000 times, then averaging over the randomisations. The white X marks the optimal parameters with largest TS value of 349. The brown, red, and white contours indicate $3, 4, 5\sigma$ from the maximum in all panels.}
    \label{fig:par_stack_psr}
\end{figure}

The results of the parameter space stacking are shown in Fig.~\ref{fig:par_stack_psr}. Each panel indicates the location of a peak TS value, and the contours are within $\Delta \mathrm{TS} = 13.2,$ 20.7, and 30.3 of the peak, 
which correspond to idealized 3, 4, and 5$\sigma$ likelihoods. 
The left panel shows the result of all 353 pulsar ROIs stacked, and a similar analysis of control field test sources is shown for comparison.

For the pulsar ROIs, the maximum TS value of 300 occurs for a flux of $1.70^{+1.41}_{-0.68} \times 10^{-10}$ \FluxUnit and a spectral index of $-0.56^{+0.56}_{-0.93}$. Using 5$\sigma$ uncertainties. The sensitivity of the LAT for these latitude ranges is  3--5$\times 10^{-10}$ \FluxUnit, about 2-3 times higher than the fluxes we are able to probe here. The spectral index is typical for 4FGL pulsars \citep{smith2019}. For the control fields, the maximum TS value of 349 occurs at a flux of $3.94^{+2.04}_{-1.22} \times 10^{-10}$ \FluxUnit and a spectral index of $-2.63^{+1.12}_{-1.38}$. At the location in parameter space where the peak TS value for the pulsars occurs, the $\Delta$(TS) = 28 between the pulsars and control fields, which corresponds naively to 4.8$\sigma$ for 2 degrees of freedom. We note that although a peak of similar TS and flux occurs in the parameter space stack for the control fields, that peak corresponds to a significantly softer spectral index than for a typical pulsar. As such, the diffuse background and/or unmodeled sources in Fermi data can be systematic effects or signal contamination in stacks that resemble the stack of the control fields. Very soft spectra generated from stacking should be treated with suspicion as our analysis and previous work \citep{paliya2020} show this may be generated when stacking blank sky. 

\subsection{Stacked Spectral Energy Distributions}
\begin{figure}
    \centering
    \includegraphics[width=\columnwidth]{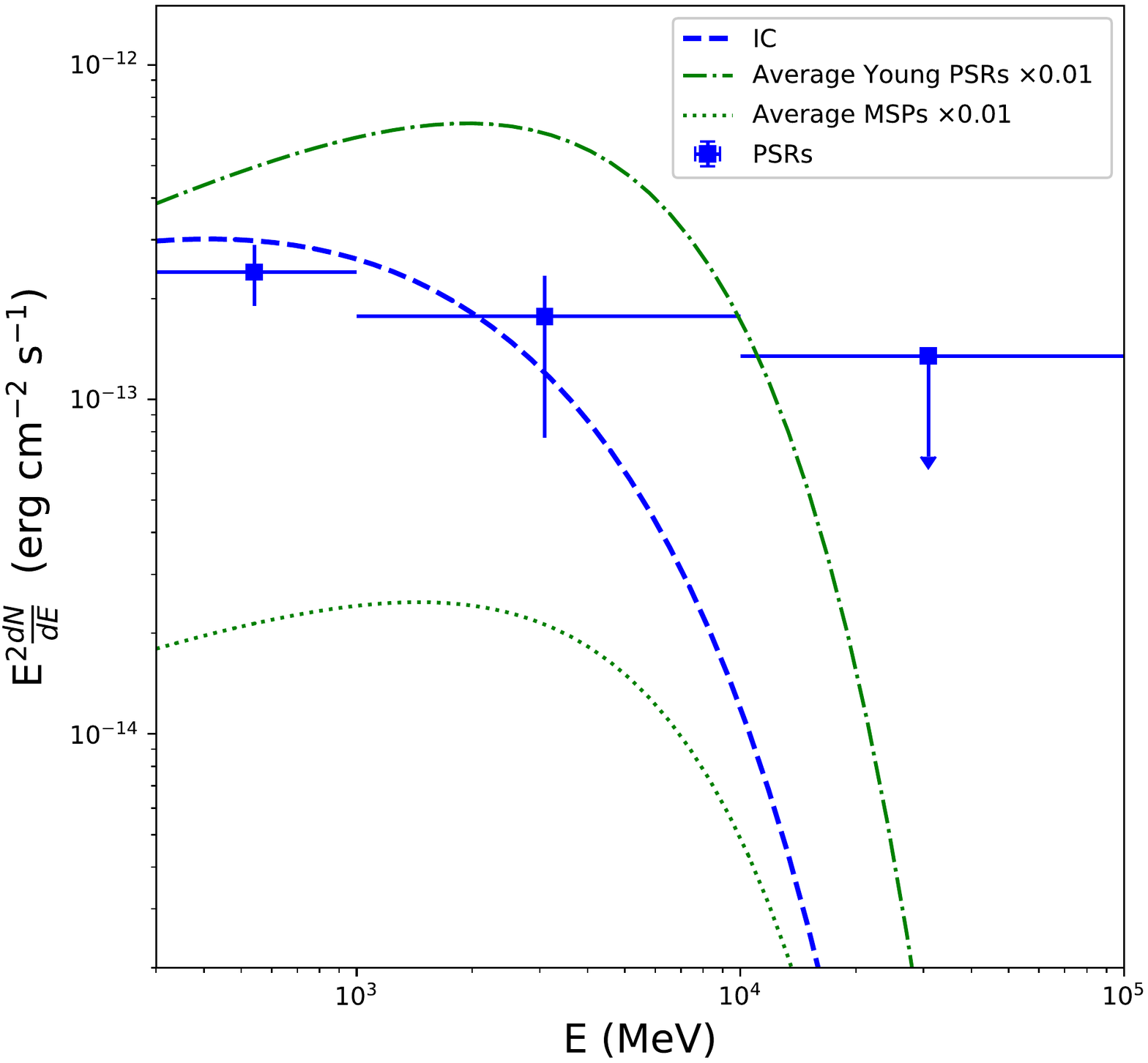}
    \caption{Stacked SED of 218 pulsars with ATNF distances that are not at 25 kpc and for which the likelihood stacking pipeline successfully converged in all three energy bins. The blue data points represent the best fit power law model in each energy bin. The blue dashed curve represents the inverse Compton scattering model result using {\tt naima} as described in \S~\ref{subsec:luminosity}. The data points and the inverse-Compton scattering model are plotted on the scale of shown on the y-axis on the left. The green dashed-dotted line and green dotted line are the average young/millisecond pulsar SEDs from \citet{mccann2015}, scaled down by 100 times to show proper comparison to the stacked SED.    }
    \label{fig:stacked_sed}
\end{figure}

We stack the signals in different photon energy bins to generate a stacked spectral energy distribution (SED) for pulsars with properly calibrated distances (Fig.~\ref{fig:stacked_sed}). To assess the SED of the collection of pulsars properly, the pulsar fluxes are recalibrated to the same distance of 1.695 kpc (the average of the distances in the sample). Here we bin the data into three energy bins: 300 MeV to 1 GeV, 1 GeV to 10 GeV, and 10 GeV to 100 GeV. While pulsars are known for their distinctively curved SEDs, in each energy bin a power law (PL) fit is used, as is standard practice\footnote{\url{https://fermipy.readthedocs.io/en/latest/advanced/sed.html}}, to simplify the analysis and reduce the degrees of freedom. The binned parameter space stack converges for 218 of the pulsar fields. We first obtain the best-fit PL index for the sample pulsars from the parameter stack. Fixing this PL index, the flux in each bin is then calculated from the best fit flux value of the largest TS in the stacked parameter space map, and uncertainties are evaluated from the $3\sigma$ spread of the flux value from the peak. The TS map of the highest energy bin does not converge in parameter space, so an upper limit is given. The stacked SED further corroborates the pulsar-like properties of the sub-threshold sources. 
To examine the effect of background contamination on the SED, the stack of pulsars without distance calibration is compared to that of the control fields. These results have similar parameter stack structures in all energy bins, although the maximum cumulative TS values vary. In particular, between 1 and 10 GeV, the parameter stack of the pulsars has a much higher peak TS than the control fields. 

To show that this stack of pulsar ROIs is consistent with previous pulsar observations, we compare the stacked SED to the previous pulsar stacking results of \citet{mccann2015}. In Fig.~\ref{fig:stacked_sed}, the green dashed-dotted line is their average best-fit spectrum for young pulsars in the catalogue with a PL index of $-1.46$ and a cutoff energy at 3.6 GeV; the green dotted curve is the average best-fit spectrum for the catalogue MSPs, with a PL index of $-1.59$ and a cutoff energy at 3.54 GeV. Our stacked SED is about an order of magnitude less luminous than the average MSP and two orders of magnitude less luminous than the average young pulsar, which shows the sensitivity of the stacking technique. To compare our results to the average SEDs of young pulsars/MSPs, the previously modeled SEDs are plotted on a different scale to match the positions of the data points. This result is consistent with the pulsars in our stacking survey being less luminous and with generally lower \Edot, which is explored further in \S~\ref{subsec:luminosity}. 

\section{Discussion}\label{sec:discussion}

\subsection{Rate of False Detection/Controlling for the Background}
\label{subsec:false}

In this study, we find that five target ROIs and two control field test sources have TS values exceeding the nominal source-detection threshold of TS $>25$ (Fig.~\ref{fig:results_1}, right panel). Modulo issues with interpreting the tails of the distribution of TS values, we can consider the control fields as effectively probing the false detection or background floor in this study. 

\begin{figure}
    \centering
    \includegraphics[width=\columnwidth]{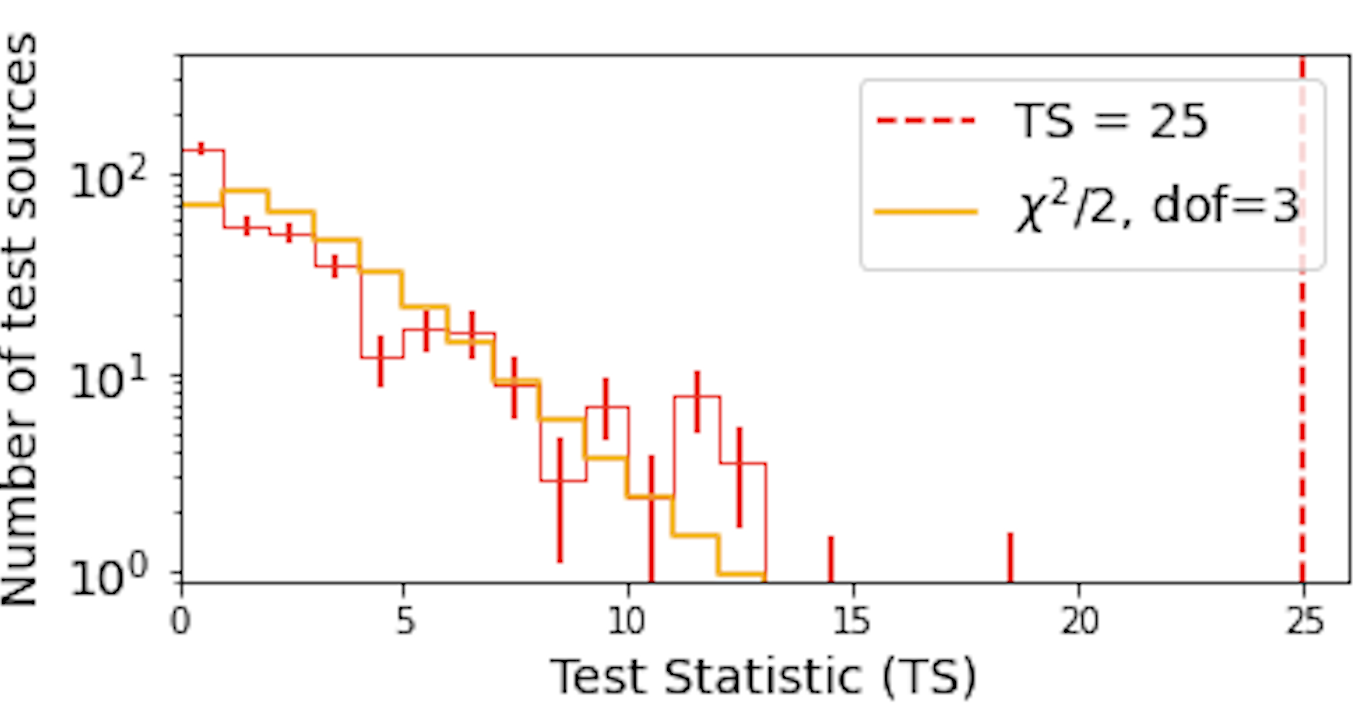}
    \includegraphics[width=\columnwidth]{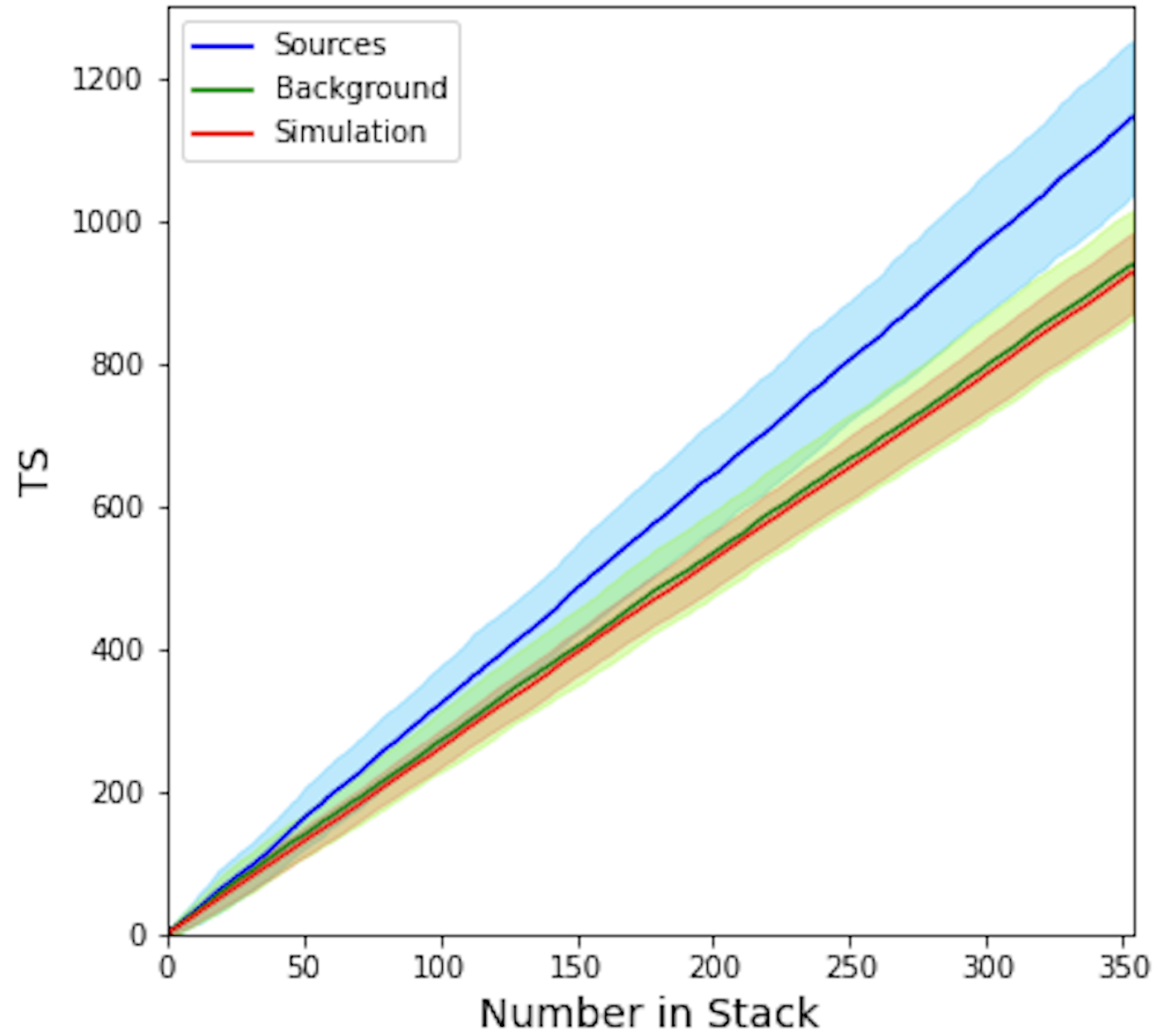}
    
    \caption{Results of ROI simulations. Top: TS distributions of null (orange histogram) and simulated test sources (red histogram). The uncertainty is estimated by randomly sampling the TS distribution of the simulated test sources. Bottom: Cumulative TS distribution of the pulsars (blue), control fields (green) and simulation (red). }
    \label{fig:sim}
\end{figure}

We also examine simulated data to understand if the control fields with high TS could be caused by uncertainties in modeling the background and/or other sources. The simulated observations are generated by the {\tt simulate\_roi} function in {\tt Fermipy}. The central source is left out of the model when creating the simulated observation, and the normalization of the isotropic background is adjusted to test the effects of varying counts noise. The simulated observations of each of the control field ROIs are then fed through the analysis pipeline with the central source added back to the model, and the TS value of the central source is measured. Without altering the background normalization, the resulting TS distribution is indistinguishable from the null,  $\chi^2/2$, which is inconsistent with the control field results in Fig.~\ref{fig:results_1} and indicates that systematic observable effects are causing a high control-field signal. TS mapping and stacking is thus highly sensitive to background modeling. An increase of only 0.1\% in the isotropic background normalization yields test source TS values that stack in a similar way as the control fields (Fig.~\ref{fig:sim}). The resulting simulated TS distribution slightly exceeds the null, but still produces no detections (TS $>25$). Thus, while spurious counts noise or background modeling systematics can account for much of the relatively large background signal we find, it seems likely that emission from unrelated sources along the line of sight contributes as well. The control field parameter stacking results  in \S~\ref{subsec:parstack_results} indicate that these contaminating or spurious sources should have notably softer spectra, however, and are distinguishable at least from pulsars as a population. 

\subsection{Luminosities}\label{subsec:luminosity}

Unsurprisingly, we find that, as a population, the undetected pulsars in our stack have significantly lower luminosities than the 4FGL-detected pulsars. This study allows us to probe the low rotation-power (\Edot) regime. Most of the pulsars in our sample also do not have \gray\ pulsations detected. 
Comparing the implied luminosities of our stack to the known \gray\ luminosities allows us to extend the known correlation between \gray\ luminosities and \Edot\ \citep{2PC}. 

During the course of this study, the Fermi-LAT collaboration updated the point source catalogue to 4FGL-DR3 containing $\sim 30$ more pulsars. For a fairer and more complete comparison, we use all the pulsars in DR3. We extracted the energy flux from 4FGL-DR3 and used the default ATNF distance modeled with YMW2016 \citep{YMW16} with a 30\%\ uncertainty to estimate their luminosities and uncertainties. We consider a beaming factor $f_\Omega = 1$, where the luminosity is the energy flux multiplied by $4\pi d^2 f_\Omega$. We estimated the luminosity of candidate pulsars based on the results from the parameter space stacking in \S~\ref{subsec:preresults}. The flux and photon index are obtained from the stacked TS maps in each \Edot\ range and integrated to get the energy flux in the 100 MeV to 100 GeV energy range. Uncertainties in energy flux are estimated from the $3\sigma$ spread of the parameters. Luminosity is then calculated by scaling to the mean of distance squared of the sub-threshold sources. The 7 pulsars without a YMW2016 distance or the 51 pulsars set to the flagged value of 25 kpc based on the dispersion measure are again excluded.

\begin{figure}
    \centering
    \hspace*{-0.5cm}
    \includegraphics[width=1.05\columnwidth]{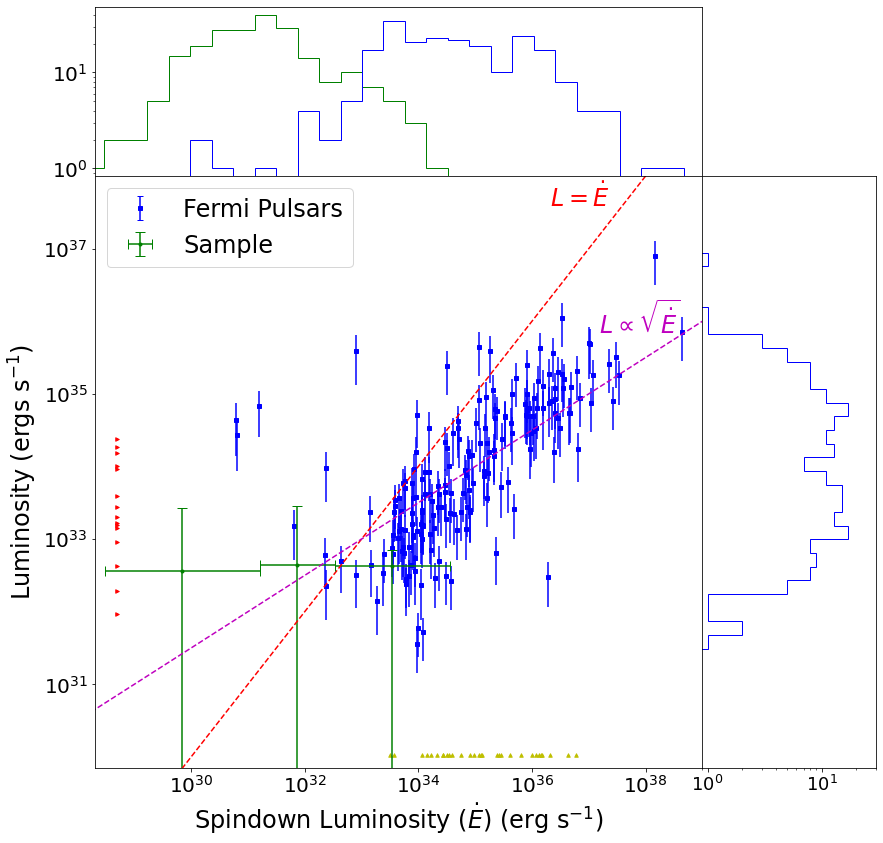}
    \caption{Gamma-ray luminosity versus \Edot, along with their distributions, for 220 4FGL-DR3 pulsars (blue) that have \Edot\ and distance measurements, and stacked candidate pulsars (green). Each \Edot\ bin for stacked pulsars, from lowest to highest, contains 104, 168, 70 pulsars, respectively. The stacking results for the candidate pulsars are indicated in green with $3\sigma$ error bars, with the luminosity estimated by multiplying the stacked energy flux with the mean of distance squared of the stacke pulsars. The red triangles on the left represent 15 4FGL pulsars without \Edot\ measurements but have a luminosity estimate. The yellow triangles at the bottom represent 30 4FGL pulsars that have \Edot\ measurements but no distance for luminosity estimation. The red dashed line indicates the maximum efficiency limit for a beaming factor $f_\Omega = 1$. The empirical dependence of $L$ on $\sqrt{\dot{E}}$ is noted by the purple dashed line. Top: \Edot\ and luminosity distributions of 4FGL-DR3 pulsars (blue) and candidate pulsars (green). Right: luminosity distributions of 4FGL-DR3 pulsars in blue.}
    \label{fig:L_Edot}
\end{figure}

The luminosity and \Edot\ of all 4FGL-DR3 pulsars and the candidate pulsars are plotted in Fig.~\ref{fig:L_Edot}. The sample pulsars appear to effectively extend the luminosity-\Edot\ dependence to low \Edot. And provide more evidence for the suggestion that at low \Edot\ the heuristic relation of $L_{\gamma}\propto \sqrt{\dot{E}}$ could change to an even stronger dependence and a very high efficiency \citep{smith2019}. 

There are two issues that might impact the results of this analysis. Firstly, pulsar distances in the YWM2016 survey are quite uncertain. These distances, when compared to some existing parallax measurements in the ATNF catalogue, can be a factor of few different, resulting in as large as an order of magnitude difference in the final luminosity estimation. Secondly, the \Pdot\ measurements of MSPs in the ATNF catalogue are not corrected to the Shklovskii effect \citep{Shklovskii1970}. Subsequently, the estimated \Edot\ of MSPs are not accurate. However, in this work, only 22 out of the sample pulsars in the stacked sample are MSPs. Especially when considering the fact that none but five of the stacked pulsars are individually detected, and none of these five pulsars are MSP, corrections for this effect are not crucial in this work.

Many models of \gray\ emission from pulsars indicate that the emission mechanism leading to the pulsed signal may turn off at low \Edot\ \citep{watters2009}. Low \Edot\ also precludes them from being pulsar wind nebula candidates. This survey was designed specifically to have clean backgrounds so the \gray\ source should originate with the neutron star itself, similar perhaps to the presumed magnetospheric off-peak emission seen in some pulsars \citep{2PC}. It is worth noticing that the majority of \gray\ pulsars are not detected with off-peak emission. This work provides better sensitivity that would likely to detect said weak off-peak emission. We investigate a simple scenario where the isotropic \gray\ emission is from an inverse Compton halo located near the light cylinder. Using the {\tt naima} code \citep{naima, naima_ics}, we assume a seed photon field from photospheric thermal emission at $10^6$ K. Integrating over all scattering angles, inverse Compton scattering from an electron distribution with a power law index of $-1.8$ and cutoff energy of 4 GeV generates a spectrum that is consistent with the fluxes and SEDs of the detected and sub-threshold sources, as shown in Fig.~\ref{fig:stacked_sed}.  The required normalization depends on the product of the non-thermal electron density and photon density (which depends on the characteristic scattering distance from the stellar photosphere). If this distance is comparable to the light cylinder radius, $\gtrsim 10$ neutron star radii, then the lepton densities required to create such a spectrum are orders of magnitude below the Goldreich-Julian density \citep{goldreich1969}. Recent development of pulsar magnetospheric models, such as the current sheet model \citep{currents_model1}, indicate the existence of such leptons at the equatorial plane of a pulsar around the light cylinder, which can provide the population for the IC scattering process.

\subsection{Further Discussion on the Pulsar \gray\ Emission Mechanism}

In this work, we make no assumptions on beaming or emission geometry of the pulsars. Out of the 362 pulsars in our sample, 105 have either no \Pdot\ or no $P$ measurement, meaning that we cannot readily search for \gray\ pulsations from them, and we cannot assume any emission geometry. The fraction of pulsars with \gray\ pulsations is a few percent to 2/3 for different \Edot\ ranges \citep{2PC}. Detectability depends not only on the pulsar luminosity and its distance from the Earth, but also the beaming geometry, e. g., the size of the beam and the viewing angle. How do we determine if the stacked signal is from faint \gray\ pulses beamed towards the Earth from a small subset of our sample, or some unknown isotropic \gray\ emission emitted by all pulsars? 

\begin{figure}
    \centering
    \includegraphics[width=\columnwidth]{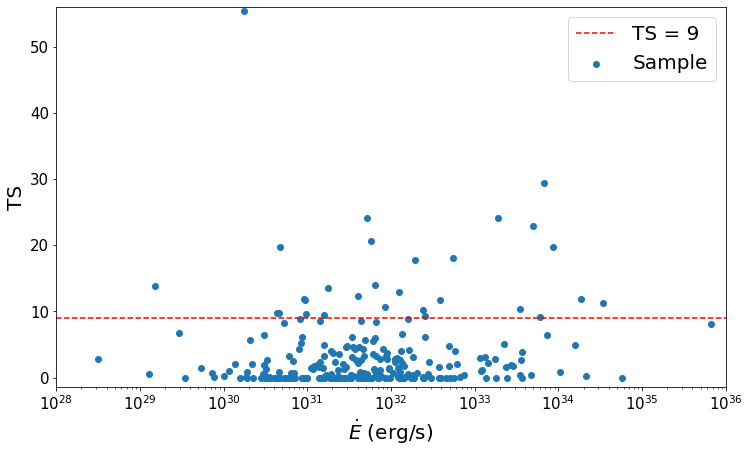}
    \caption{TS vs. \Edot\ of our sample pulsars. Pulsars without \Edot\ measurements are plotted as green triangles on the left. }
    \label{fig:TS_edot}
\end{figure}

We first examine the correlation between pulsar \Edot\ and their TS values from the single ROI binned likelihood analysis. As shown in Fig.~\ref{fig:TS_edot}, there is no obvious correlation between TS and pulsar \Edot . Since pulsed \gray\ luminosity is generally dependent on \Edot , the stacked signal is not dominated by just a few high \Edot\ pulsars. Instead these pulsars may have weak, unpulsed or very broadly pulsed gamma-ray emission that originates, for example, from the equatorial region around the light cylinder of the pulsar, or is more isotropic. 

\begin{figure*}
    \centering
    \includegraphics[width=0.25\textwidth]{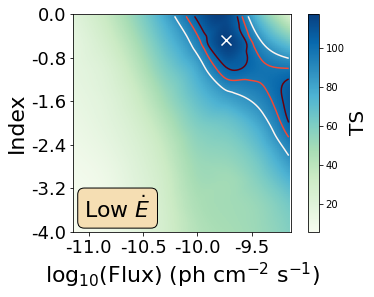}
    \includegraphics[width=0.25\textwidth]{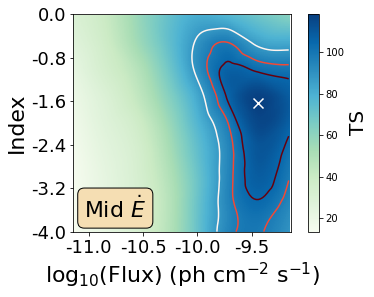}
    \includegraphics[width=0.25\textwidth]{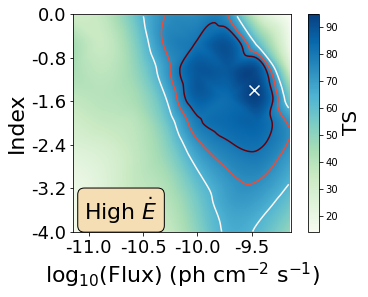}\\
    \includegraphics[width=0.45\textwidth]{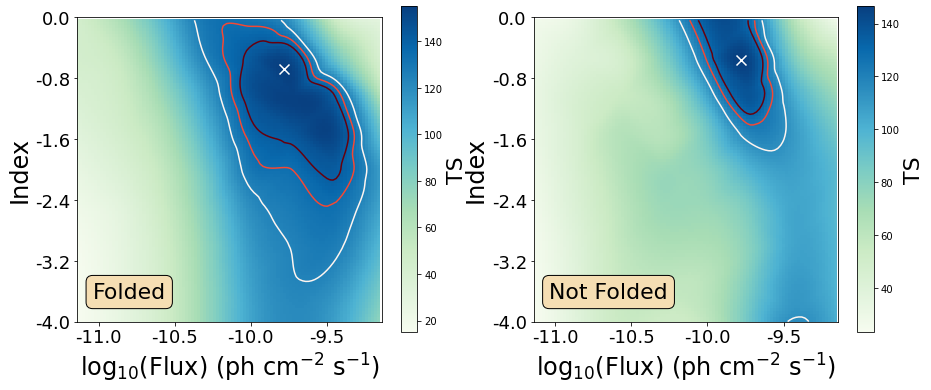}
    \includegraphics[width=0.45\textwidth]{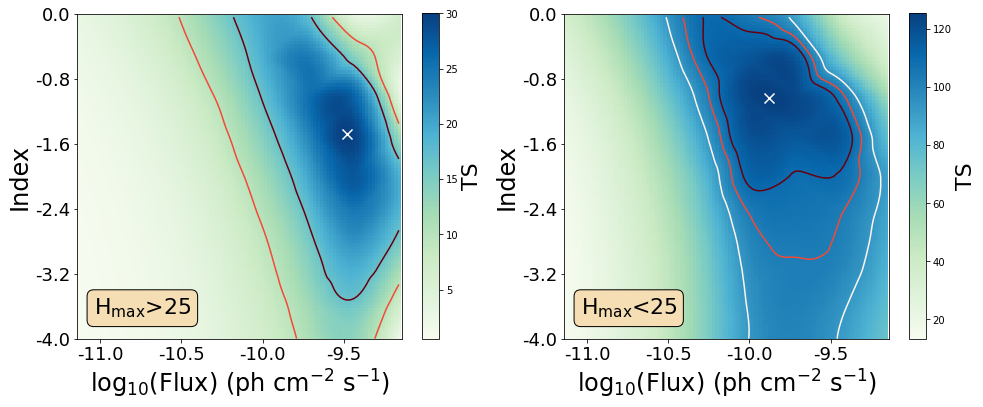}
    
    \caption{Top panels: Parameter space stacking results of the three different \Edot\ bins in Fig.~\ref{fig:L_Edot}. Bottom left and bottom mid-left: parameter space stacking results of the two sub-populations of sample pulsars that have been folded in \gray s (left) or not (right). Bottom mid-right and bottom right: parameter space stacking results of the two sub-populations of sample pulsars with H$_{\rm max} > 25$ (left) or H$_{\rm max} < 25$ (right).}
    \label{fig:ps_edot}
\end{figure*}

Similarly, there is no significant difference in the parameter space stacking results after grouping by \Edot\ (Fig.~\ref{fig:ps_edot}). The peak TS values in each \Edot\ bin are very similar, and their parameter uncertainties overlap considerably. This result reinforces that the \gray\ emission mechanism from the three distinctive pulsar populations might be very similar. Given the large error bars on the luminosity estimation of these populations, we cannot say with any further certainty that this mechanism is \Edot\ dependent or not. We defer the exploration of low \Edot\ pulsar \gray\ emission mechanism to future study.

Previous searches for \gray\ pulsations from pulsars \citep[e.g., ][]{smith2019} did not detect pulsations from a large fraction of high \Edot\ pulsars. This result might be dominated by the emission geometry rather than their \gray\ luminosities. From our sample of pulsars, 174 were phase-folded in \citet{smith2019}. Here, we try to correlate our results with the $H_{\rm max}$ values introduced in \citet{smith2019}. $H_{\rm max}$ is defined as the maximum H-test value when H-test is performed in six different ways (refer to Sec. 3 in \citet{smith2019} for further details). The highest $H_{\rm max}$ was 1212 for J0514-4408 which is a 4FGL-DR3 recorded pulsar, an indication of a significantly detected periodic signal. We note that 5 pulsars have $H_{\rm max} >25$, including J0922+0638, 148 have $H_{\rm max}<25$, and 21 did not return a valid $H_{\rm max}$. Again there is no significant difference between the parameter stacks of the subgroups of pulsars that were phase-folded or not, or resulted in high or low $H_{\rm max}$ values. The stack of the 5 pulsars with $H_{\rm max}>25$ seems to exhibit a significant peak with TS = 30 as shown in the bottom left panel of Fig.~\ref{fig:ps_edot}. It is worth noting that the parameter space stack of two of these five pulsars did not converge, and the parameter space stack of J0922+0638 as shown in Fig.~\ref{fig:J0922_validation} dominates this stack. The stack for the 148 low $H_{\rm max}$ pulsars shows a peak with TS = 120. Randomly selecting 5 pulsars out of this list indicates a small chance ($< 2\%$) of a peak TS larger than 30. While the parameters are not well constrained in either case, this result seems to imply a correlation between the peak TS and the $H_{\rm max}$ values of the pulsars being stacked, but with a very small number of high $H_{\rm max}$ pulsars, the correlation is not significant. One can also search for the existence of weak, isotropic \gray\ emission from pulsars by examining the off-peak phases of detected pulsars, which is reserved for future study.

\section{Conclusions}

In this work we examined the \gray\ emission from 362 locations that coincide with ATNF pulsars using Fermi-LAT data. Our sample ranges widely from young to old pulsars, magnetars, MSPs, and low \Edot\ pulsars. 

The stacked signal from the 362 target sources exceeds that of the background. Using the bootstrap range as a very conservative measure of the uncertainty indicates at least a $2.5\sigma$ difference. Stacking in spectral parameter space implies a higher significance as well as a clear distinction between the target and background SEDs.
Stacking the TS profiles of the candidate pulsars in spectral parameter space (flux and photon spectral index) indicates a pulsar-like spectrum assuming a cutoff energy at 823 MeV, with an index of $-0.56^{+0.56}_{-0.93}$, which is consistent with known pulsar SEDs. The characteristic \gray\ flux between 300 MeV to 100 GeV flux of $1.70^{+1.41}_{-0.68} \times 10^{-10}$ \FluxUnit\ is about a factor of two lower than the Fermi-LAT point source sensitivity. The stacked SED of these pulsars is also consistent with previous population studies. The candidate pulsars luminosities roughly follow the expected dependence of $L_{\gamma}$ on $\dot{E}$, extrapolated to low spin-down power.

Five of the candidate pulsars, J0038-2501, J0922+0638, J1610-17, J1705-04 and J2336-01, have TS values exceeding the detection threshold of 25, with J0922+0638 detected in 4FGL-DR3. Details about these five candidate detections are discussed in Appendix~\ref{appendix:discussion}. 

\section*{Acknowledgements}

The authors would like to express gratitude towards the reviewer, D. Smith, for his constructive feedback and comments on this work. This work was supported in part by NSF grants AST-1831412, AST-2219090, and AST-1852360, as well as PSC-CUNY award \#63785-00 51, and the CUNY Research Scholars Program. The corresponding author is also supported by the Australian Research Council Centre of Excellence for Gravitational Wave Discovery (OzGrav), through project number CE170100004. The authors thank D. Zurek, M. Bailes and S. Stevenson for very useful discussions. This project made use of computational systems and network services at the American Museum of Natural History supported by the National Science Foundation via Campus Cyberinfrastructure Grant Awards \# 1827153 (CC$*$ Networking Infrastructure: High Performance Research Data Infrastructure at the American Museum of Natural History).

\section*{Data Availability}

Fermi-LAT photon data are available through Fermi-LAT data server.
Fermi-LAT analysis results (the output .xml or .npy files from Fermipy) of target pulsars and control fields can be shared upon request to the first author.
Post-processing analysis scripts will be shared upon request to the first author, given that a proper citation to this paper is provided in the work from the people that make the request.


\appendix

\section{Validating Parameter Space Analysis Method}
\label{subsec:validation}
As a proof of concept, we performed the same parameter stack scheme on J0922+0638 because it was not recorded in 4FGL-DR2, but was recorded in 4FGL-DR3 as 4FGL J0922.3+0644. DR3 gives a PL model of the source, with TS = 29, a power law spectral index of $-2.65 \pm 0.14$, and a flux from 1 to 100 GeV to of $(1.19 \pm 0.28) \times 10^{-10}$ \FluxUnit , which can be converted to $(8.2 \pm 3.0) \times 10^{-10}$ \FluxUnit\ for photons between 0.3 to 100 GeV. 

\begin{figure}
\hspace{-2cm}
\includegraphics[width=1.4\columnwidth]{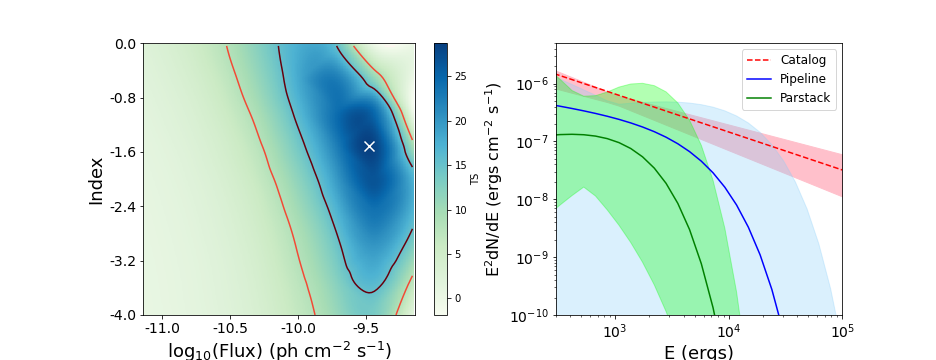}
    \caption{Left: TS map of PSR J0922+0638 (4FGL J0922.3+0644) in parameter space, following the parameter stacking analysis scheme described in \S~\ref{subsec:stacking_method}. The white X marks the peak TS value of 28 in the chosen region in parameter space and the black and red contours represent the $3\sigma$ and $5\sigma$ away from the peak. Right: Comparison of the SED of J0922+0638 generated using the catalogue parameters (red) as described in text, results from the standard binned likelihood analysis as recorded in Table~\ref{tab:detections} (blue), and parameter stack results from the left panel (green). The shaded areas each represent the 95\% uncertainty in the flux estimation. }
    \label{fig:J0922_validation}
\end{figure}

The parameter space analysis of the pulsar in the left panel of Fig.~\ref{fig:J0922_validation}, yields a peak TS = 28 with a \gray\ flux (300 MeV - 100 GeV) of $(3.0^{+2.5}_{1.4}) \times10^{-10}$ \FluxUnit , consistent with the catalogue value, as well as a power law index of $-1.6^{+1.6}_{-1.7}$ and a fixed cutoff energy of 823 MeV. We plotted the SED of the pulsar using the results of the spectral parameter space TS map, using the parameters from the catalogue as described in the last paragraph, and using the results from a standard binned likelihood analysis as tabulated in Table.~\ref{tab:detections} (Fig.~\ref{fig:J0922_validation}). These results are all consistent with each other within the uncertainties. 

\section{Determination of Latitude Cut}\label{appendix:latcut}

In the main text, only the 362 pulsars that are 20\degr\ away from the plane are used in the stacks, which was motivated by analysing the impact of different latitude cuts on the stack sensitivity. This appendix describes the determination of this latitude limit, which derived from an analysis all pulsars located 5\degr\ away from the Galactic plane not already detected in \gray s. For sources within 20\degr\ of the Galactic plane, only those 30\degr\ away in longitude from the Galactic center were analyzed. A total of 726 pulsars satisfy these selection criteria.

\begin{figure}
    \centering
    \includegraphics[width=\columnwidth]{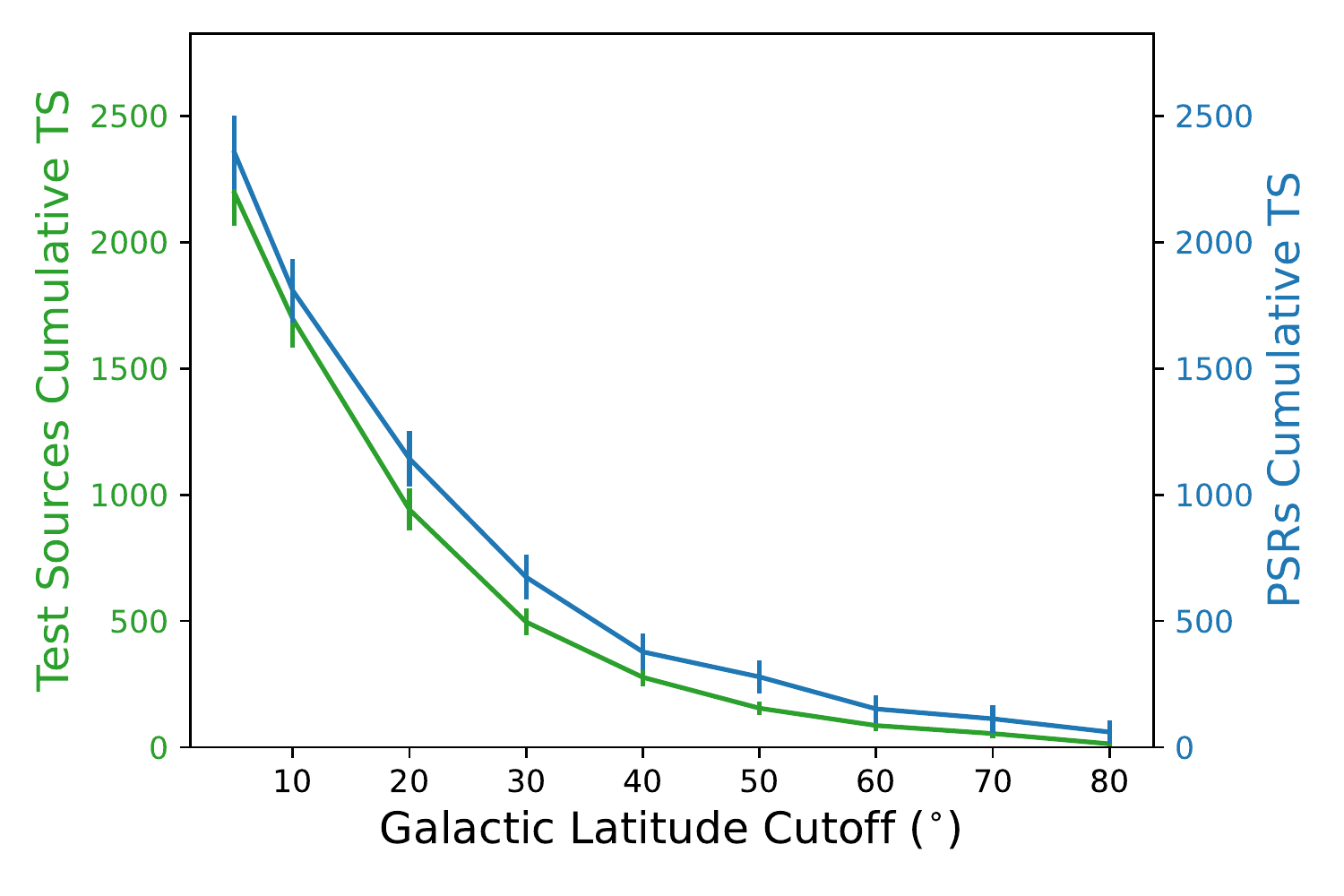}
    \caption{Cumulative TS values and uncertainties of the pulsars (blue line with error bars) and the corresponding control field test sources (green dashed line with error bars) with the same spatial distribution, plotted against the minimum latitude selection.}
    \label{fig:lat_cut}
\end{figure}

Fig.~\ref{fig:lat_cut} shows the final cumulative TS value for the pulsars and control field test sources as a function of the lower latitude cut. The error bars on both curves are the standard deviation of the final cumulative TS values using the bootstrap technique described in \S~\ref{subsec:stacking_method}. The sensitivity (separation of the stacks) is best given a 20\degr\ or 30\degr\ latitude cut. Overall, as expected, when the latitude cut is low, extending the source population into the Galactic plane, the noise and false positive rate increase. When the latitude cut is higher, the noise impact is small but the number of sources decreases quickly. We chose the 20\degr\ latitude cut in this study to maximize both sensitivity and the number of possible detected pulsars. 

\section{Discussion of Pulsar Candidates with TS $>25$}\label{appendix:discussion}

As mentioned above, five pulsars from our sample have TS $>$ 25, the usual detection threshold. To further assess the \gray\ emission from these locations, we found the peak TS value each of these pulsars using the {\tt localize} routine in {\tt fermipy}. We also reoptimized the ROI. Their optimized locations and uncertainties, original locations and uncertainties from the ATNF, and SEDs, are shown in Fig.~\ref{fig:SEDs}. Their spectral properties obtained in this work are shown in Table~\ref{tab:detections}. The same analysis was performed on the two control field test sources that have TS $> 25$ and the results are also shown in Fig.~\ref{fig:SEDs} and Table~\ref{tab:detections}. One candidate, PSR J0922+0638, has now been subsequently reported in the 4FGL-DR3 as a newly detected \gray\ pulsar. 

PSR J0038-2501 and PSR J1705-04, have well fitted parameters and pulsar-like SEDs. PSR J1610-17, has a poorly fit index, and its cutoff energy is extremely low, which overall results in a very soft SED. PSR J2336-01 has a relatively soft spectral index, a largely unconstrained cutoff energy, and its SED is slightly more consistent with a simple power law spectrum rather than a PLEC shape. The curvature significance of each source is quantified by the TS$_{\rm curv}$ value included in Table~\ref{tab:detections}. 

We also examined the individual map for these five pulsars. J0922, J0038 and J1705 all have relatively good individual maps in parameter space. The other two, J1610 and J2336 have upper-limit/background like maps in parameter space. 

\begin{figure*}
    \includegraphics[width=0.45\textwidth]{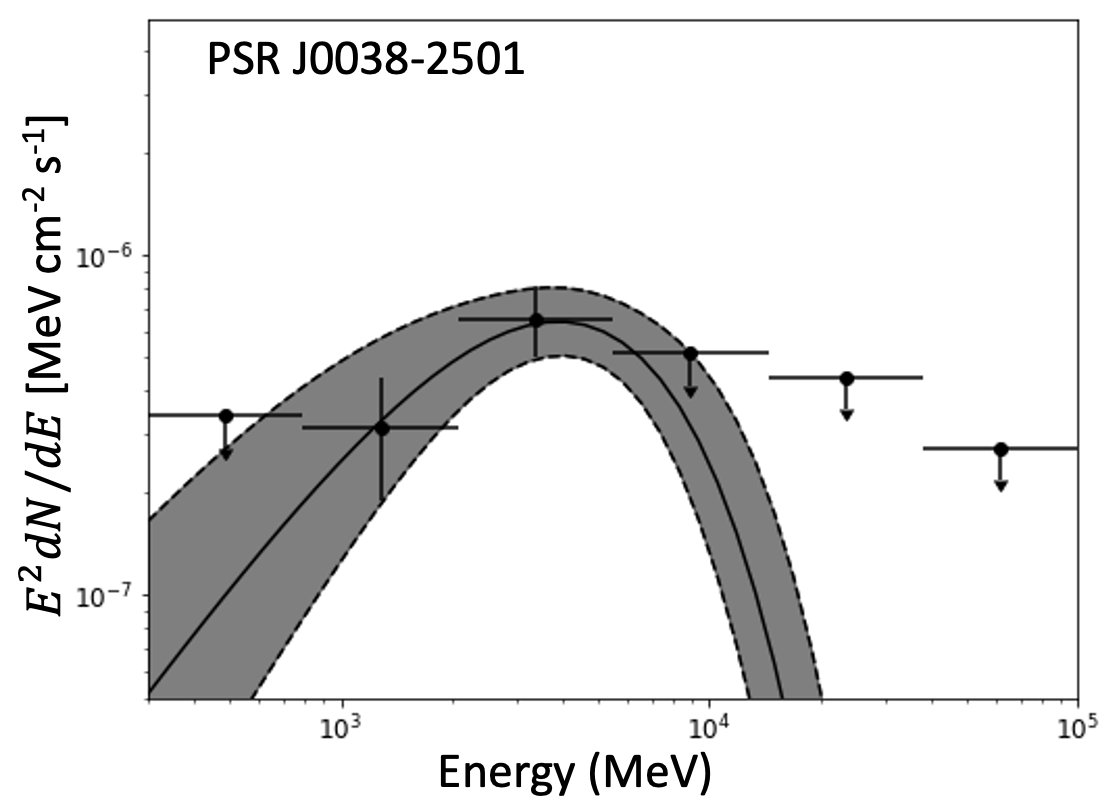}
    \includegraphics[width=0.45\textwidth]{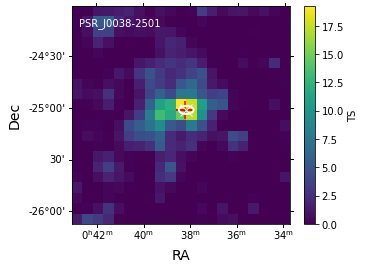}\\
    \includegraphics[width=0.45\textwidth]{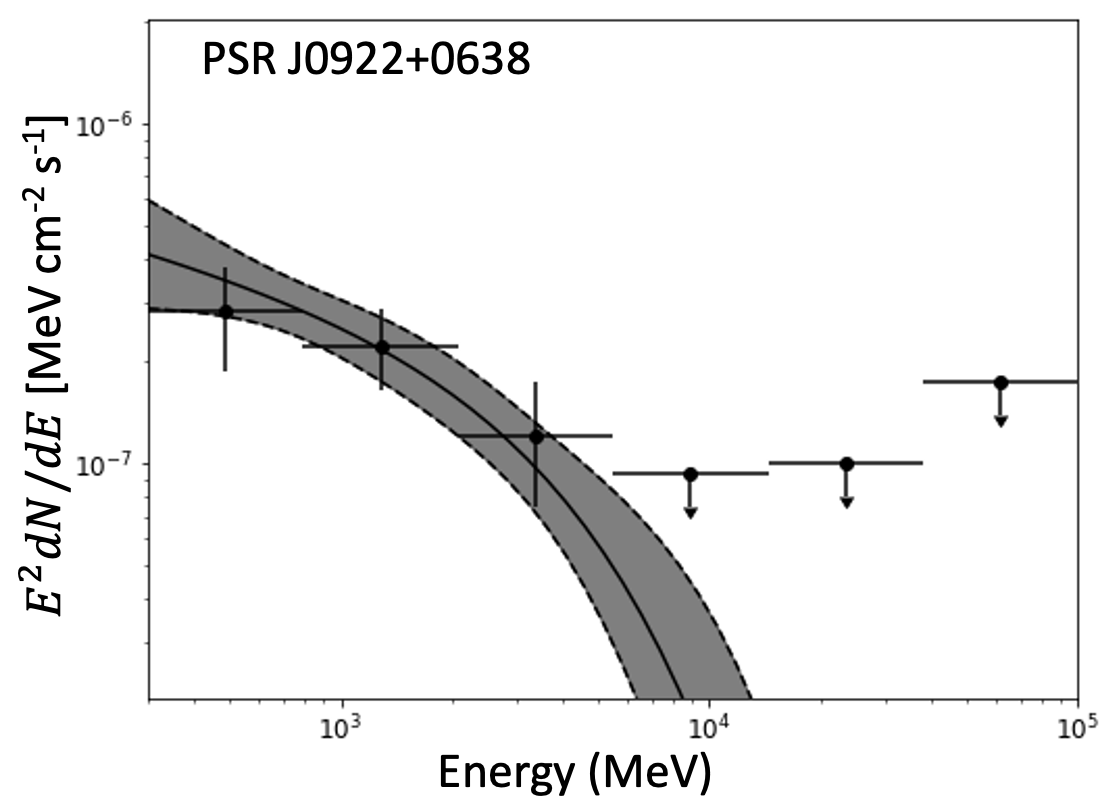}
    \includegraphics[width=0.45\textwidth]{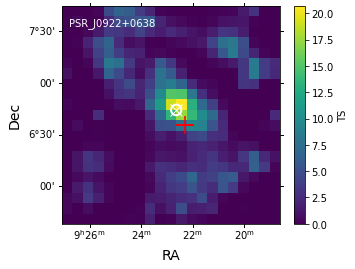}\\
    \includegraphics[width=0.45\textwidth]{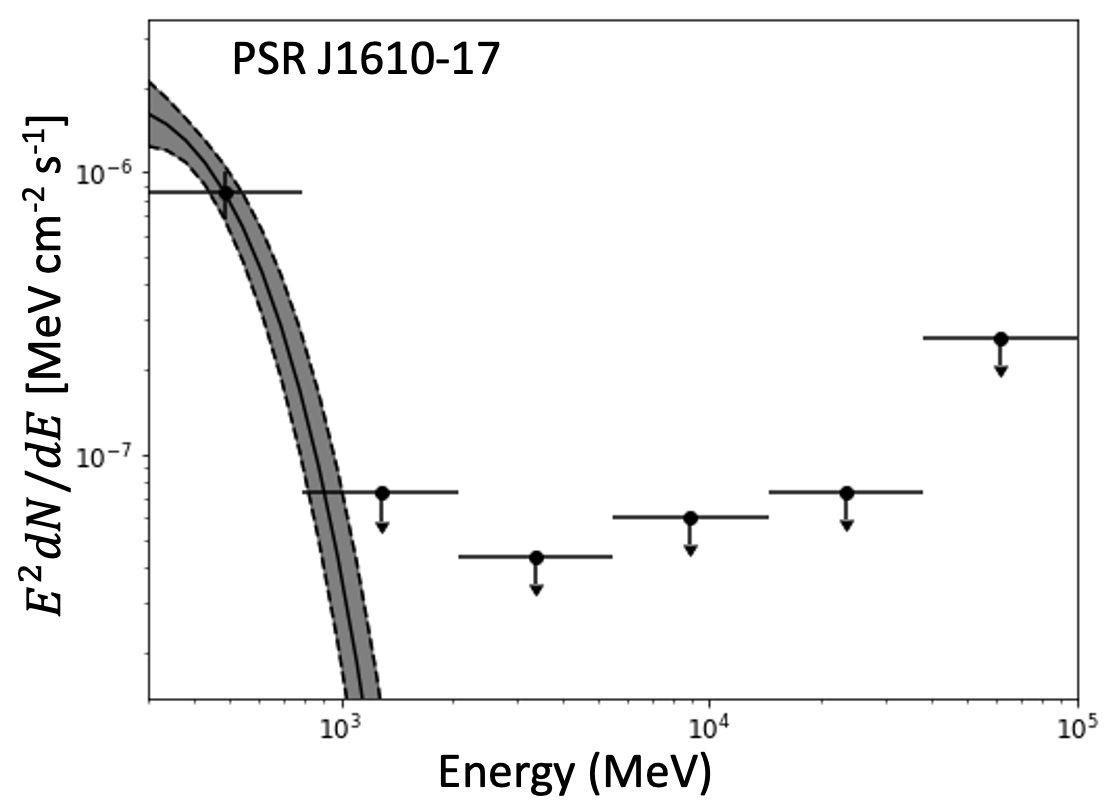}
    \includegraphics[width=0.45\textwidth]{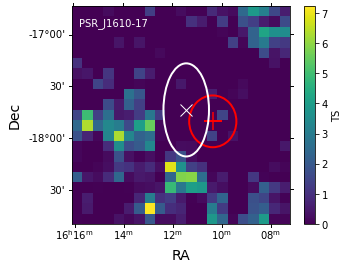}\\
    \includegraphics[width=0.45\textwidth]{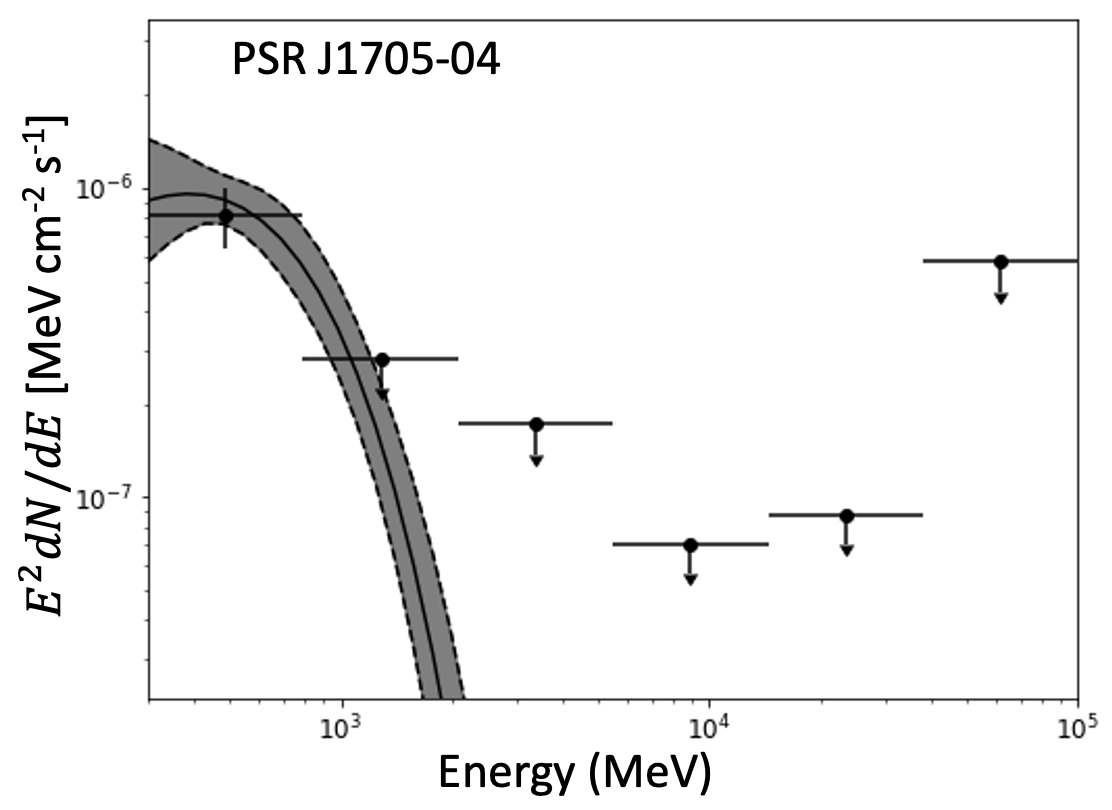}
    \includegraphics[width=0.45\textwidth]{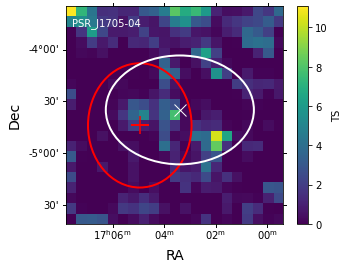}\\
    
\end{figure*}
\begin{figure*}
\ContinuedFloat
    \includegraphics[width=0.45\textwidth]{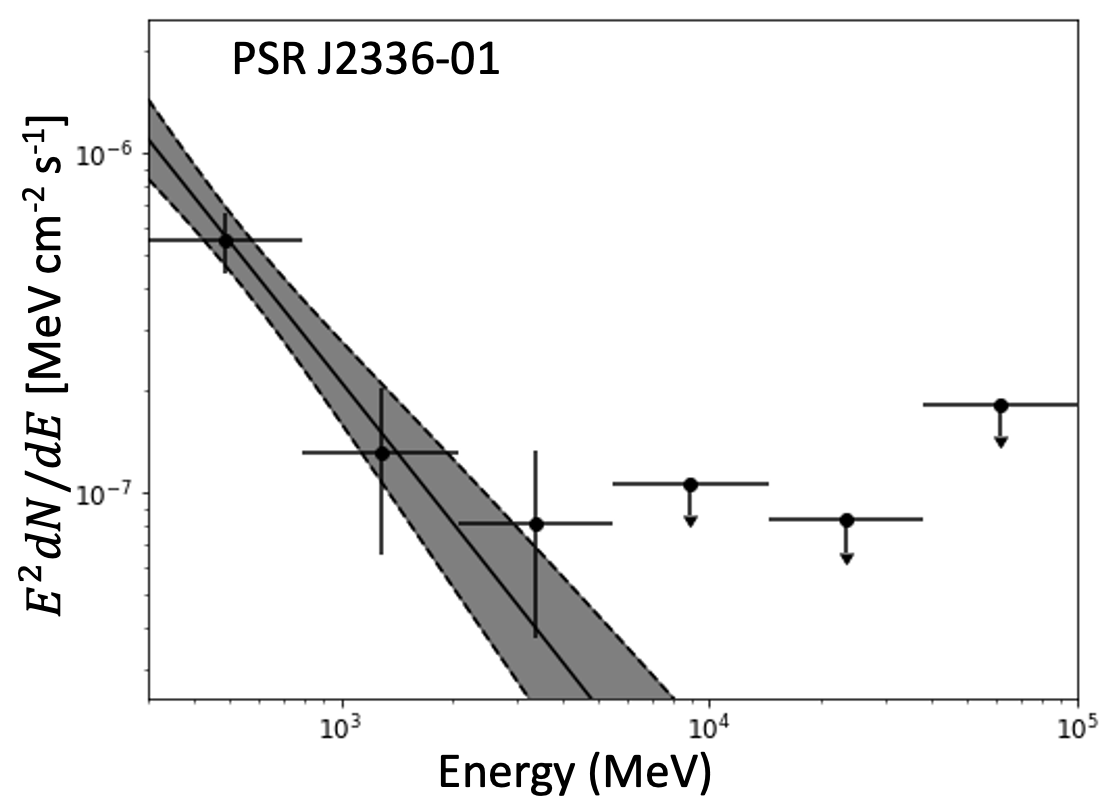}
    \includegraphics[width=0.45\textwidth]{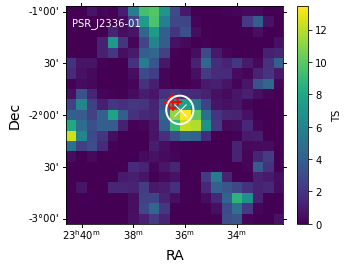}\\
    \includegraphics[width=0.45\textwidth]{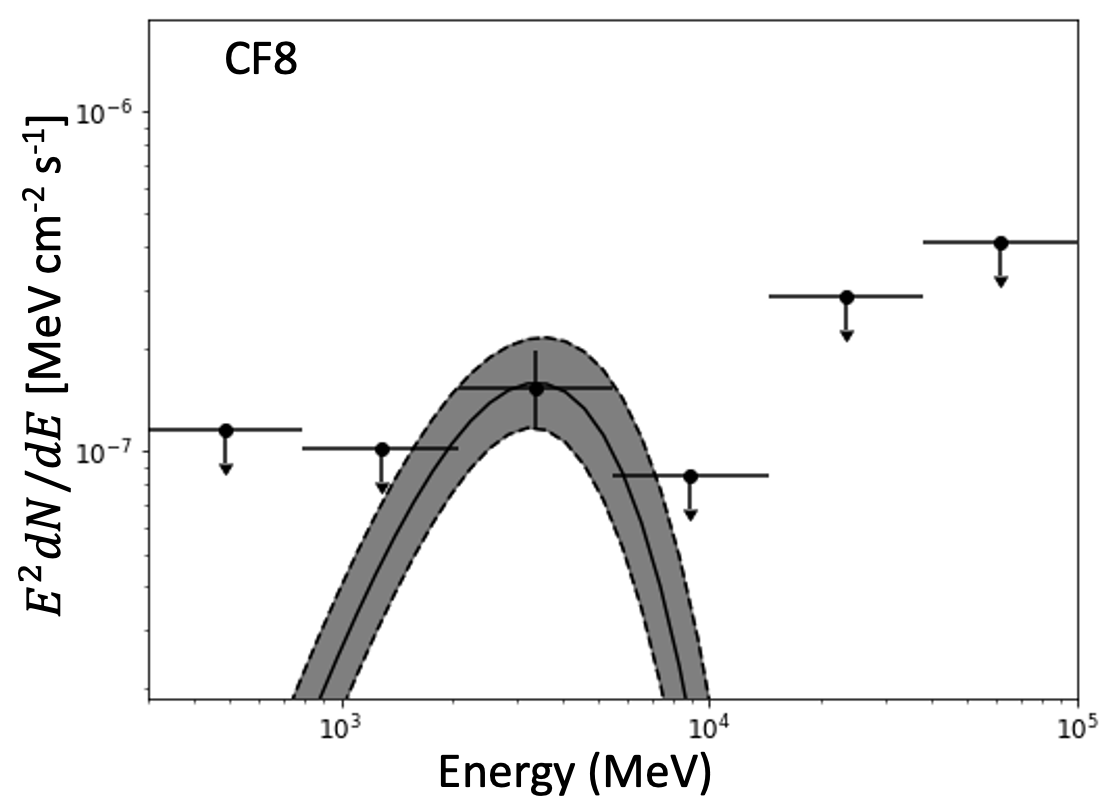}
    \includegraphics[width=0.45\textwidth]{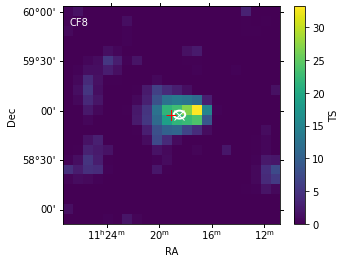}\\
    \includegraphics[width=0.45\textwidth]{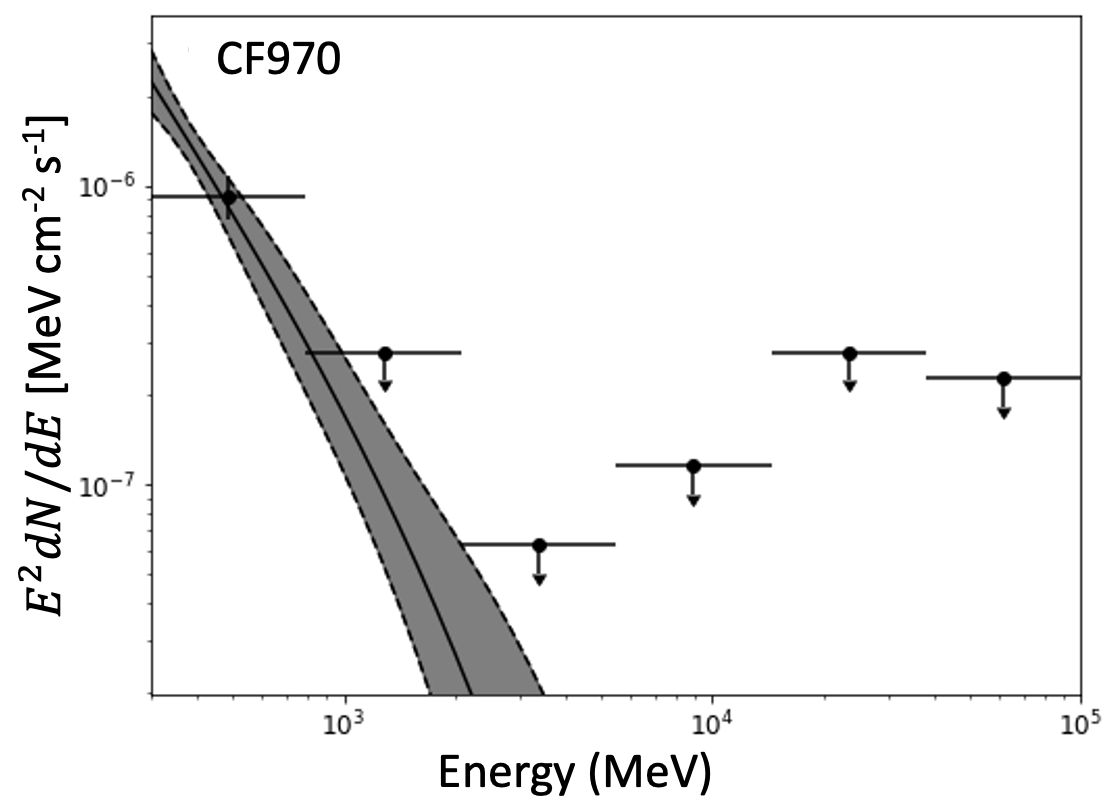}
    \includegraphics[width=0.45\textwidth]{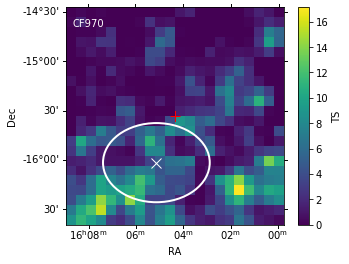}\\
    \caption{Spectral energy distributions and TS maps of target pulsars and control field test sources with TS $>25$. Fluxes with errorbars are shown in each energy bin with TS $>4$, and the rest are shown with 95\%\ upper limits. We also plot the best-fit PLEC (black solid line) spectral models from the likelihood analysis over the full energy range along with the uncertainties (gray shaded area), with the parameters given in Table~\ref{tab:detections}. The red + and ellipse represent the radio position and uncertainty of each pulsar; the white x and ellipse indicate the optimal \gray\ location and 95\% uncertainty.}
    \label{fig:SEDs}
\end{figure*}

Overall, PSR J1705-04 is potentially a good candidate for follow-up study to see whether its \gray\ flux can be confirmed to be associated with the pulsar. PSR J0038-2501 has a TS value as high as 55 before localization and reoptimization, the TS value is 33 after localization and reoptimization. It is only 0.04\degr\ away from 4FGL J0038.2-2459, a \gray-detected flat spectrum radio quasar, making it hard to distinguish the signal from the two sources. PSR J2336-01 and PSR J1610-17 seem more likely to be spurious or due to flux from an unassociated source near the line of sight given their soft and/or low curvature SEDs. Additionally, PSR J2336-01 is within 1\degr\ of three FSRQs in the catalogue: 4FGL J2335.4-0128, 4FGL J2338.0-0230, and 4FGL J2333.4-0133. However, the likelihood analysis pipeline could separate these signals. While phase-folded pulse detection is needed to confirm the pulsar nature of these candidates, attempting to establish \gray\ timing solutions is reserved for future work.

Two of these five pulsars have \Edot\ measurements. Their efficiencies, $L_\gamma / \dot{E}$ = 53 and 334, respectively, which are extraordinarily large. These extremely high over-unity efficiencies could be an indicator that these gamma-rays do not originate from the pulsars. However, like all the other sample pulsars, the distances of these two are largely uncertain, a factor which impacts the estimated luminosity greatly. Additionally, both of these pulsars have \Edot\ much lower than the deathline value of $10^{33}$ erg/s. Given the limited understanding of pulsar behaviours in this regime, they also could possibly have $f_{\Omega}$ values very different from unity due to different emission geometries.

It might be worth noting that, another pulsar in our list, J1455-3330, that have slightly higher, yet still undetected signals ($10 < $TS$ < 25$), is also now present in the 4FGL-DR3 catalogue.

\begin{table*}
\begin{adjustwidth}{-0.8cm}{}
\caption{Candidate Pulsar and Control Field Detections\label{tab:detections}}

\resizebox{1.05\textwidth}{!}{
\begin{tabular}[width=\textwidth]{l|cccccccccccc}
\hline
  & RA (Fermi) & Dec (Fermi) & RA (radio) & Dec (radio) & Photon Flux & Index &  Cutoff Energy & TS & TS$_{\rm curv}^{\rm a}$ & Luminosity & \Edot\ &  Dist\\
 & (h m s)  & ($\degr\ ^\prime$) & (h m) & ($\degr\ ^\prime$) & ($\times 10^{-9}$ ph cm$^{-2}$ s$^{-1}$) &   & GeV & & & ($\times 10^{32}$ ergs s$^{-1}$) & ($\times 10^{30}$ ergs s$^{-1}$) & (kpc)\\
\hline
J0038-2501 & $00^{\mathrm h}38^{\mathrm m}7.\!\!^{\mathrm s}51(17.\!\!^{\mathrm s}05)$ & $-25\degr01.\mkern-4mu^\prime32(2. \mkern-4mu^\prime33)$ & $00^{\mathrm h}38.\!\!^{\mathrm m}17(0.^{\mathrm s}16)^{\rm b}$ & $-25\degr01.\mkern-4mu^\prime51(. \mkern-4mu^\prime03)^{\rm b}$ & $0.68 \pm 0.22$ & $-0.48 \pm 0.80$ & $2.52 \pm 1.33$ & 33.0 & $^{h}$ & $0.94 \pm 0.60$ & 1.77 & 0.604 \\

J0922+0638$^{c}$ & $09^{\mathrm h}22^{\mathrm m}35.\!\!^{\mathrm s}44(11.\!\!^{\mathrm s}95)$ & $06\degr47.\mkern-4mu^\prime21(3. \mkern-4mu^\prime06)$ & $09^{\mathrm h}22.\!\!^{\mathrm m}23(0.\!\!^{\mathrm m}06)^{\rm d}$ & $06\degr38.\mkern-4mu^\prime39(0. \mkern-4mu^\prime28)^{\rm d}$ & $0.96 \pm 0.23$ & $-2.28 \pm 0.57$ & $3.86 \pm 5.05$ & 38.9 & 2.2 & $3.67 \pm 2.31$ & $1.1$ & $6.8$ \\

J1610-17 & $16^{\mathrm h}11^{\mathrm m}15.\!\!^{\mathrm s}62(52.\!\!^{\mathrm s}78)$ &  $-17\degr43.\mkern-4mu^\prime36(26. \mkern-4mu^\prime9)$ & $16^{\mathrm h}10.\!\!^{\mathrm m}18(0.\!\!^{\mathrm m}95)^{\rm e}$ & $-17\degr50^\prime(0. \mkern-4mu^\prime25)^{\rm e}$ & $2.04 \pm 0.38$ & $0.16 \pm 1.64$ & $0.11 \pm 0.04$ & 30.3 & 2.5 & $23.07 \pm 14.57$ &  & 3.768\\ 

J1705-04 & $17^{\mathrm h}03.\!\!^{\mathrm m}447(2.\!\!^{\mathrm m}860)$ & $-04\degr32.\mkern-4mu^\prime01(31. \mkern-4mu^\prime54)$ & $17^{\mathrm h}05^{\mathrm m}(2^{\mathrm m})^{f}$& $-04\degr41^\prime(36^\prime)^{f}$ &$2.03 \pm 0.44$ & $-0.36 \pm 2.10$ & $0.23 \pm 0.17$  &  26.9 & 5.65 & $ 0.09\pm 0.05$ &  & 0.21 \\

J2336-01 & $23^{\mathrm h}36^{\mathrm m}19.\!\!^{\mathrm s}84(31.\!\!^{\mathrm s}57)$ & $-01\degr 55.\mkern-4mu^\prime43(8.\mkern-4mu^\prime25)$ & $23^{\mathrm h}36.\!\!^{\mathrm m}6(0.\!\!^{\mathrm m}2)^{g}$ & $-01\degr51^\prime(3^\prime)^{g}$ & $1.48 \pm 0.28$ & $-3.41 \pm 0.42$ & $107.3 \pm 659.4$ &  33.8 & 1.8 & $8.29 \pm 5.22$ &  & 2.393 \\
\hline
CF8 & $11^{\mathrm h}18^{\mathrm m}26.\!\!^{\mathrm s}87(13.\!\!^{\mathrm s}87)$ & $58\degr 58.\mkern-4mu^\prime68(2.\mkern-4mu^\prime41)$  &  &  & $0.09 \pm 0.03$ & $1.49 \pm 0.002$ & $0.97 \pm 0.18$ & 30.8 & 6.3 & $1.19 \pm 0.82$ &  & \\

CF970 & $16^{\mathrm h}05.\!\!^{\mathrm m}133(2.\!\!^{\mathrm m}160)$ & $-16\degr 1.\mkern-4mu^\prime80(24.\mkern-4mu^\prime03)$ &  &  & $2.46 \pm 0.39$ & $-3.77 \pm 1.51$ & $1.64 \pm 5.96$ & 42.0  &  & $5.91 \pm 3.71$ &  & \\

\hline
\multicolumn{10}{p{0.95\linewidth}}{$^{a}$ TS$_{\rm curv}$ = TS$_{\rm PLEC} -$ TS$_{\rm PL}$}\\
\multicolumn{10}{p{0.95\linewidth}}{$^{b}$ \citet{acd+19}}\\
\multicolumn{10}{p{0.95\linewidth}}{$^{c}$ Reported in 4FGL-DR3}\\
\multicolumn{10}{p{0.95\linewidth}}{$^{d}$ \citet{hlk+04}}\\
\multicolumn{10}{p{0.95\linewidth}}{$^{e}$ \citet{bb10}}\\
\multicolumn{10}{p{0.95\linewidth}}{$^{f}$ \citet{karako2015}}\\
\multicolumn{10}{p{0.95\linewidth}}{$^{g}$ \citet{sanidas2019}}\\
\multicolumn{10}{p{0.95\linewidth}}{$^{h}$ Power law model did not converge, indicating that this model is strongly disfavored.}\\
\end{tabular}
}
\end{adjustwidth}
\end{table*}

For completeness, we note that there are also two TS $>25$ control field test sources found in our library. Performing a similar analysis on them resulted in poorly localized positions and uncertain spectral parameters. The fitting results did not converge for the control field labeled as CF970. CF8, while seemingly well-localized and with a pulsar-like SED, has a suspiciously low index uncertainty, which usually indicates a poor fit. Thus, we consider this control field detection to be spurious as well.

\section{Stacking Results from 100 MeV}\label{appendix:100MeV}

\begin{figure*}
\includegraphics[width=0.9\textwidth]{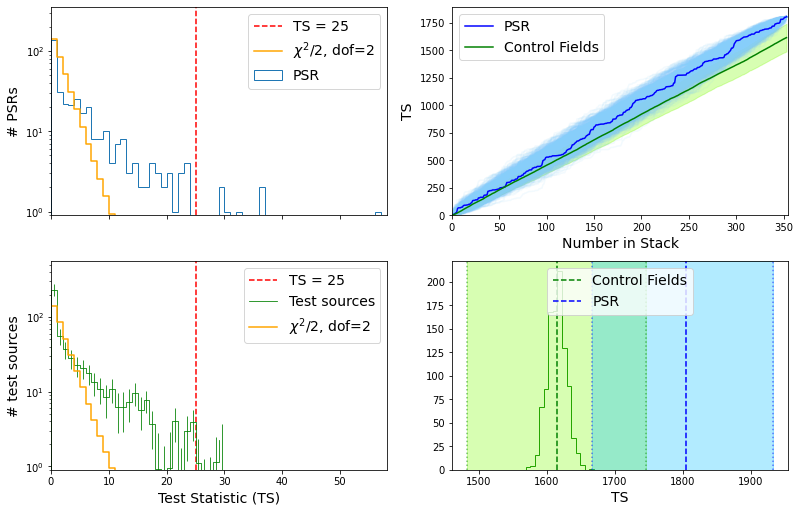}\\
\includegraphics[width=\textwidth]{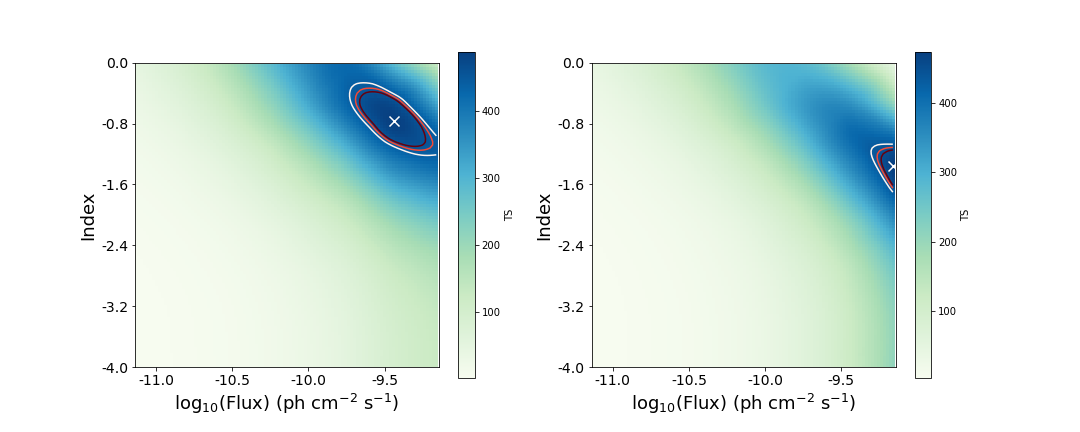}

    \caption{Top left: TS$_{100}$ values of target pulsars using data from 100 MeV to 100 GeV data. Bottom left: TS values of control field test sources from analysis with 100 MeV to 100 GeV data. 
    The orange histogram represents the $\chi^2$/2 distribution for 2 degrees of freedom. The red dashed vertical line represents TS = 25, the presumed detection threshold. Right: Cumulative TS$_{100}$ values of the target pulsars (blue) and control field test sources (green). The light blue curves represent stacks after one randomly re-ordering of the TS distribution of the pulsars. The light green shaded areas represent uncertainties of the stack for control fields estimated from bootstrap resampling of the stacks.
    Bottom left: Parameter space stack of 315 pulsars. The white X marks the optimal parameters with the largest TS value of 397, and the contour shows the 5$\sigma$ range. Bottom right: Parameter stack of 315 control fields. }
    \label{fig:ts100_dist}
\end{figure*}

Recent work has revealed that due to the worse resolution and contamination from the backgrounds, parameter stacking that includes photons below 300 MeV may introduce a spurious population of extremely soft sources \citep{paliya2020}. Here, we test this claim by comparing the stacking analysis results using the 300 MeV to 100 GeV data to those using the 100 MeV to 100 GeV data following the same prescription outlined in \S~\ref{sec:obs}. In Fig.~\ref{fig:ts100_dist}, we plot the TS$_{100}$ (TS value of a source obtained using data between 100 MeV and 100 GeV) distribution and the cumulative TS$_{100}$ distribution of both the pulsar ROIs and control field ROIs. While the stacked pulsar signal still persists when examining the cumulative TS$_{100}$ distributions, the difference from the control stack is less significant and the standard deviation of the mean final cumulative TS value is larger.

We also investigated the 100 MeV to 100 GeV pulsar and control field ROIs using the parameter space stacking technique. Only 315 of the pulsar ROIs converged at all points in parameter space using this method. In Fig.~\ref{fig:ts100_dist}, we show the parameter stacking results of these 315 pulsar and control field ROIs that converged between 100 MeV and 100 GeV. The control field stack is again performed by randomly choosing 315 fields out of the 540, repeating this 1000 times, and averaging. The parameter space stacking of pulsars shows no significant difference from the results using 300 MeV to 100 GeV data in Fig.~\ref{fig:par_stack_psr}. The control fields stacking results, still have a softer spectral index than the pulsars, though slightly harder than the stack results between 300 MeV and 100 GeV. The average flux is also higher, enough to breach the detection threshold of Fermi and contribute significant spurious signal.

Overall, stacking the control fields from 100 MeV does seem to capture the noise/background contamination in the low energy range. Hence, we recommend using the higher energy range, due to the following additional considerations: 1) In our analysis, we used a curved PLEC spectrum for our sources instead of a simple power law. A larger systematic effect could occur for power law sources. 2) Above 300 MeV, the improved resolution of Fermi-LAT can greatly aid in localizing faint signals and in making less ambiguous source associations. 3) We also saw a decrease in the control field background signal when excluding the lower energies.

\section{Three-Dimensional Parameter Space Stack}\label{appendix:3dparstack}

With three parameters ({\tt prefactor}, {\tt index1} and {\tt cutoff} energy) for PLEC fitting, a complete parameter-space stack could include, in addition to the flux and index, the cutoff energy. It is a feature of the known population of \gray\ pulsars that the cutoff energies lie within a very small range, which is 1.7 GeV for MSPs, and 0.9 GeV for radio-loud \gray\ pulsars \citep{dr3}. It is therefore reasonable to set the cutoff energy to a fixed value in our methods. 

\begin{figure*}
    \centering
\includegraphics[width=0.7\textwidth]{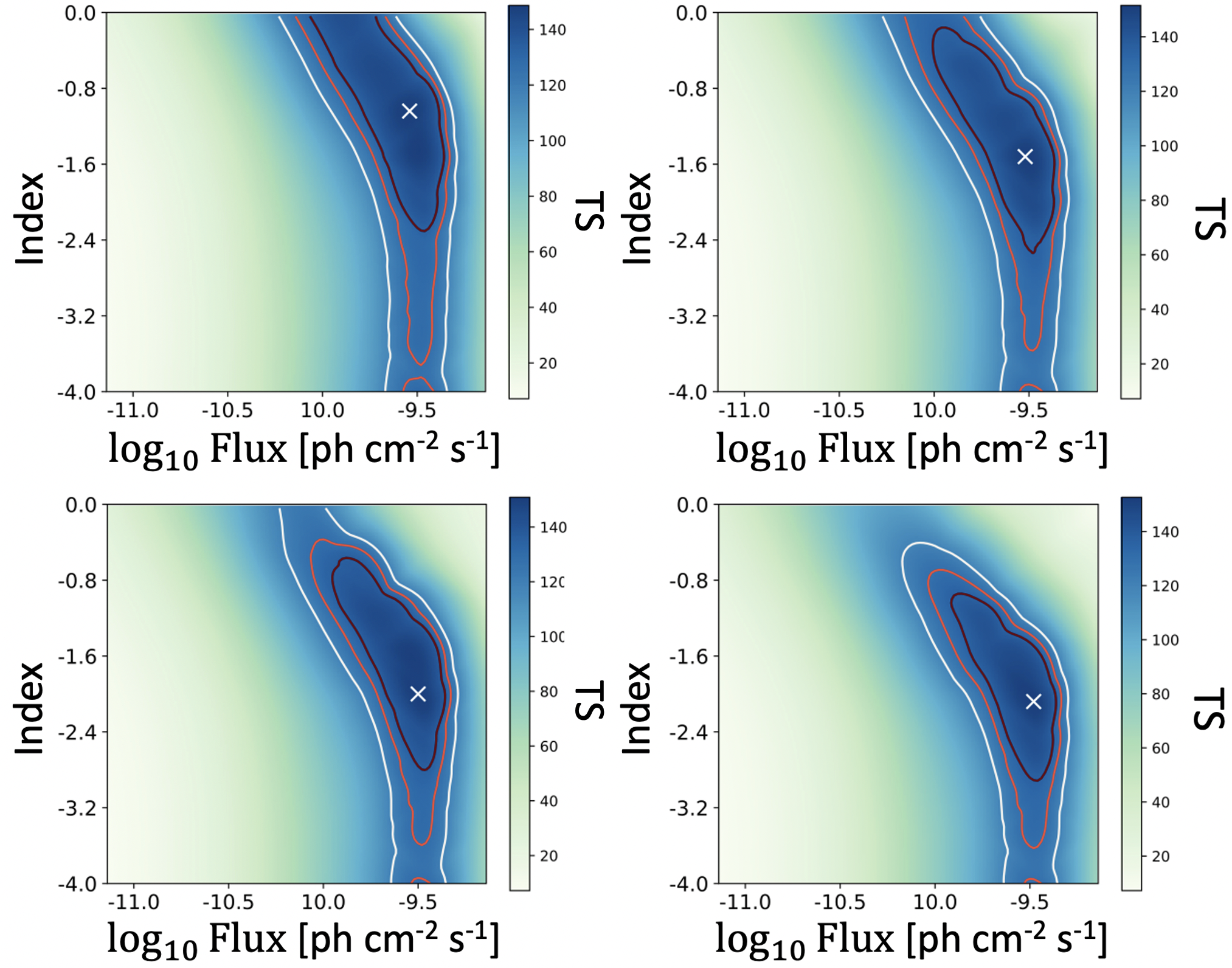}
 \caption{Parameter space stacking of 138 randomly sampled pulsar ROIs with 4 different cutoff energies of 655 MeV, 823 MeV, 1034 MeV and 1300 MeV, from top left to top right to bottom left to bottomr right, respectively. }
    \label{fig:3d_parstack}
\end{figure*}

To further test the dependence on cutoff energy, we calculated the same parameter stack using data from 300 MeV to 100 GeV at 4 different cutoff energies with a sample selection of 138 pulsars from the target list. The results for cutoff energies of 655 MeV, 823 MeV, 1034 MeV and 1300 MeV, are shown in Fig.~\ref{fig:3d_parstack}. There is a slight softening of the spectrum with increasing cutoff energy, which is consistent with the trends for 4FGL pulsars and the analytical relation between the two quantities derivable directly from the PLEC model. However, the $5\sigma$ range is large compared with the trend, and the peak TS value does not heavily depend on the choice of the cutoff energy. We therefore chose to fix the cutoff energy at the median value of all the PLEC 4FGL pulsars for the analysis presented in \S~\ref{subsec:preresults}, which also conveniently reduces the degrees of freedom.


\bsp	
\label{lastpage}
\end{CJK*}
\end{document}